\documentclass[12pt,epsf]{article}

\usepackage{amsmath}
\usepackage{amssymb}

\allowdisplaybreaks

\begin{document}

\baselineskip=18.8pt plus 0.2pt minus 0.1pt

\makeatletter

\@addtoreset{equation}{section}
\renewcommand{\theequation}{\thesection.\arabic{equation}}
\renewcommand{\thefootnote}{\fnsymbol{footnote}}
\newcommand{\beq}{\begin{equation}}
\newcommand{\eeq}{\end{equation}}
\newcommand{\bea}{\begin{eqnarray}}
\newcommand{\eea}{\end{eqnarray}}
\newcommand{\nn}{\nonumber\\}
\newcommand{\hs}[1]{\hspace{#1}}
\newcommand{\vs}[1]{\vspace{#1}}
\newcommand{\Half}{\frac{1}{2}}
\newcommand{\p}{\partial}
\newcommand{\ol}{\overline}
\newcommand{\wt}[1]{\widetilde{#1}}
\newcommand{\ap}{\alpha'}
\newcommand{\bra}[1]{\left\langle  #1 \right\vert }
\newcommand{\ket}[1]{\left\vert #1 \right\rangle }
\newcommand{\vev}[1]{\left\langle  #1 \right\rangle }

\newcommand{\ul}[1]{\underline{#1}}
\newcommand{\Slash}[1]{{\ooalign{\hfil/\hfil\crcr$#1$}}}

\makeatother

\begin{titlepage}
\title{
\vspace{1cm}
A Master Action for $D=11$ Supergravity in the Component Formulation
}
\author{Yoji Michishita
\thanks{
{\tt michishita@edu.kagoshima-u.ac.jp}
}
\\[7pt]
{\it Department of Physics, Faculty of Education, Kagoshima University}\\
{\it Kagoshima, 890-0065, Japan}
}

\date{\normalsize June, 2016}
\maketitle
\thispagestyle{empty}

\begin{abstract}
\normalsize
We give a solution to the classical master equation of $D=11$ supergravity
in the conventional component formulation.
Based on a careful investigation of the symmetry algebra including terms 
proportional to the equation of motion,
we construct an explicit expression of the master action in an order-by-order manner.
\end{abstract}
\end{titlepage}

\section{Introduction}

Supersymmetric field theories often show tractableness of quantum behavior.
One of them is the cancellation of loop divergences.
Especially this property has been studied in the most supersymmetric theory
i.e. $D=11$ supergravity and its dimensional reductions.
A partial list of literature on recent discussions on this subject is \cite{d10}-\cite{ls15}.

Since perturbative calculation of supergravity amplitudes is quite complicated,
the on-shell condition for external states is often imposed to simplify the calculation
and to use various methods for circumventing the complexity of direct calculation.
If we want to perform covariant off-shell calculation of loop amplitudes 
in the standard way based on Feynman rules,
we need to introduce ghosts and fix the local symmetry.
Because the local symmetry of $D=11$ supergravity is reducible and 
its algebra is open, Faddeev-Popov gauge fixing procedure does not work,
and we have to use field-antifield formalism, or Batalin-Vilkovisky (BV) formalism
(For reviews, see e.g. \cite{h90}-\cite{gps94}).
In this formalism additional fields are introduced and using them 
it is necessary to construct a master action which satisfies the classical master equation. 
For $D=11$ supergravity, this has been done in the pure spinor superfield formulation in
\cite{c0910}. In this formulation the master action takes very simple form. 
However, as a cost of the simplification, superfields contain huge number of 
auxiliary fields, and the relation between them and the conventional component
expression is not immediately clear.

In this paper we give a master action for $D=11$ supergravity 
in conventional component expression.
Although introduction of ghosts makes the Feynman rules even more complicated and
it may not be practical to use the action for computing amplitudes,
it expresses the relation between structure functions of the symmetry algebra
compactly, and it may be useful for formal arguments about properties of amplitudes.
After quickly reviewing $D=11$ supergravity and fixing the notation in Section 2 and Appendix A,
we introduce ghosts and investigate Jacobi identity in Section 3.
Then we construct a master action in Section 4. Section 5 contains a conclusion.
As is usual in supergravity theories, we need tedious calculation, especially for Fierz transformation.
Such calculations are made with the help of symbolic manipulation program Mathematica and
the package for gamma-matrix algebra GAMMA\cite{g01}.
We give the outlines of the calculation in Appendix B, C, and D.

After this work was finished, we realized that \cite{cds09} has already 
studied component expression of the field-antifield formulation of $D=11$ supergravity.

\section{$D=11$ supergravity and its local symmetries}

The action of $D=11$ supergravity 
$S_0=\frac{1}{2\kappa^2}{\cal S}_0$, which consists of the vielbein $e_\mu{}^a$,
the gravitino $\psi_\mu^\alpha$, and the 3-form $A_{\mu\nu\lambda}$, is given by
\bea
{\cal S}_0 & = & \int d^{11}xe\Big[
R(\omega)-\frac{i}{2}\bar{\psi}_\mu\Gamma^{\mu\nu\lambda}D_\nu\Big(\frac{1}{2}(\omega+\hat{\omega})\Big)\psi_\lambda
-\frac{1}{2\cdot 4!}F_{\mu_1\dots\mu_4}F^{\mu_1\dots\mu_4}
\nn & &
-\frac{i}{192}(\bar{\psi}_{\nu_1}\Gamma^{\nu_1\mu_1\dots\mu_4\nu_2}\psi_{\nu_2}
 +12\bar{\psi}^{\mu_1}\Gamma^{\mu_2\mu_3}\psi^{\mu_4})\cdot
 \frac{1}{2}(\hat{F}_{\mu_1\dots\mu_4}+F_{\mu_1\dots\mu_4})
\nn & &
 +\frac{\sigma}{(144)^2}\epsilon^{\mu_1\dots\mu_{11}}A_{\mu_1\mu_2\mu_3}F_{\mu_4\dots\mu_7}F_{\mu_8\dots\mu_{11}}
\Big],
\label{sol0}
\eea
where
\bea
\omega_{\mu ab} & = & \omega_{\mu ab}(e)+\frac{1}{2}(T_{\mu ab}-T_{ab\mu}-T_{b\mu a}),
\nn
e_\nu{}^ae_\lambda{}^b\omega_{\mu ab}(e) & = &
 e_{\nu a}\p_{[\lambda}e_{\mu]}{}^a+e_{\lambda a}\p_{[\mu}e_{\nu]}{}^a-e_{\mu a}\p_{[\nu}e_{\lambda]}{}^a,
\nn
T^\mu{}_{ab} & = & \frac{i}{4}\bar{\psi}_{[a}\Gamma^\mu\psi_{b]}
 -\frac{i}{8}\bar{\psi}_{\nu}\Gamma_{ab}{}^{\mu\nu\lambda}\psi_\lambda,
\nn
F_{\mu_1\dots\mu_4} & = & 4\p_{[\mu_1}A_{\mu_2\mu_3\mu_4]},
\nn
\hat{\omega}_{\mu ab} & = & \omega_{\mu ab}
 -\frac{i}{16}\bar{\psi}_{\nu}\Gamma_{ab\mu}{}^{\nu\lambda}\psi_\lambda,
\nn
\hat{F}_{\mu_1\dots\mu_4} & = & F_{\mu_1\dots\mu_4}
 +\frac{3}{2}i\bar{\psi}_{[\mu_1}\Gamma_{\mu_2\mu_3}\psi_{\mu_4]},
\eea
and $\sigma=\pm 1$ is the sign factor defining the 10th gamma matrix (see \eqref{sgng10}).
$T^\mu{}_{ab}$ is the torsion determined by 1.5 order formalism. For our notation about spinors
see Appendix A.

This action has four local symmetries: the supersymmetry, the diffeomorphism, 
the local Lorentz symmetry, and the 3-form gauge symmetry.
The supersymmetry transformation $\delta^S$ is given by
\begin{align}
\delta^S_\xi e_\mu{}^a & = \frac{i}{4}\bar{\xi}\Gamma^a\psi_\mu, & 
\delta^S_\xi e_a{}^\mu & = -\frac{i}{4}\bar{\xi}\Gamma^\mu\psi_a,
\label{symse} \\
\delta^S_\xi \psi_\mu & = \wt{D}_\mu\xi, & 
\delta^S_\xi\bar{\psi}_\mu & = \wt{D}_\mu\bar{\xi},
\label{symspsi} \\
\delta^S_\xi A_{\mu\nu\lambda} & = -\frac{3}{4}i\bar{\xi}\Gamma_{[\mu\nu}\psi_{\lambda]}, & &
\label{symsa}
\end{align}
where the parameter $\xi$ is a Majorana spinor, and
\bea
\wt{D}_\mu\xi & := & D_\mu(\hat{\omega})\xi
 +\frac{1}{288}(\Gamma^{\mu_1\dots\mu_4}{}_\mu
 -8\delta_\mu{}^{\mu_1}\Gamma^{\mu_2\mu_3\mu_4})\xi\hat{F}_{\mu_1\dots\mu_4},
\nn
\wt{D}_\mu\bar{\xi} & := & D_\mu(\hat{\omega})\bar{\xi}
 -\frac{1}{288}\bar{\xi}(\Gamma^{\mu_1\dots\mu_4}{}_\mu
 +8\delta_\mu{}^{\mu_1}\Gamma^{\mu_2\mu_3\mu_4})\hat{F}_{\mu_1\dots\mu_4}.
\eea
The diffeomorphism $\delta^D$, the local Lorentz transformation $\delta^L$, and 
the 3-form gauge transformation $\delta^A$ take the standard form:
\bea
\delta^D_\epsilon e_\mu{}^a & = & -\epsilon^\nu\p_\nu e_\mu{}^a-\p_\mu\epsilon^\nu e_\nu{}^a, \nn
\delta^D_\epsilon \psi_\mu^\alpha & = & -\epsilon^\nu\p_\nu \psi_\mu^\alpha
 -\p_\mu\epsilon^\nu \psi_\nu^\alpha, \nn
\delta^D_\epsilon A_{\mu\nu\lambda} & = & -\epsilon^\rho\p_\rho A_{\mu\nu\lambda}
 -3\p_{[\mu}\epsilon^\rho A_{|\rho|\nu\lambda]},
\label{symd}
\eea
\beq
\delta^L_\lambda e_\mu{}^a = \lambda^a{}_be_\mu{}^b,\quad
\delta^L_\lambda\psi_\mu = \frac{1}{4}\lambda_{ab}\Gamma^{ab}\psi_\mu,\quad
\delta^L_\lambda A_{\mu\nu\lambda} = 0,
\label{syml}
\eeq
\beq
\delta^A_\theta e_\mu{}^a = 0,\quad
\delta^A_\theta \psi_\mu = 0,\quad
\delta^A_\theta A_{\mu\nu\lambda} = 3\p_{[\mu}\theta_{\nu\lambda]}.
\label{syma}
\eeq
Hatted fields $\hat{\omega}_{\mu ab}$ and $\hat{F}_{\mu\nu\lambda\rho}$
are supercovariant i.e. their supersymmetry transformation do not contain
derivatives of the parameter:
\bea
\delta^S_\xi\hat{\omega}_{\mu ab} & = & \frac{i}{4}e_a{}^\nu e_b{}^\lambda\Big[
\bar{\xi}\Gamma_{[\nu}\wt{D}_{\lambda]}\psi_\mu
+\bar{\xi}\Gamma_{[\lambda}\wt{D}_{|\mu|}\psi_{\nu]}
-\bar{\xi}\Gamma_\mu\wt{D}_{[\nu}\psi_{\lambda]}
\nn & &
-\frac{1}{144}\bar{\xi}(
 \Gamma_{\nu\lambda}{}^{\mu_1\dots\mu_4}
 +24\delta_{[\nu}{}^{\mu_1}\delta_{\lambda]}{}^{\mu_2}\Gamma^{\mu_3\mu_4}
)\psi_\mu\hat{F}_{\mu_1\dots\mu_4}
\Big],
\\
\delta^S_\xi\hat{F}_{\mu\nu\lambda\rho} & = & 
 -3i\bar{\xi}\Gamma_{[\mu\nu}\wt{D}_\lambda\psi_{\rho]}
 -i\bar{\xi}\Gamma^\sigma\psi_{[\mu}\hat{F}_{\nu\lambda\rho]\sigma}.
\eea
Note that in the above expressions all the $O(\psi^3)$ terms are hidden in $\wt{D}_\mu$
and $\hat{F}_{\mu_1\dots\mu_4}$.
To show that the explicit $O(\psi^3)$ terms are canceled for $\delta^S_\xi\hat{F}_{\mu\nu\lambda\rho}$
we need Fierz identity \eqref{fierz1}.

Taking variation of the action with respect to $\psi_\mu$, we obtain
\beq
\delta{\cal S}_0 = \int d^{11}x
 (-ie)\delta\bar{\psi}_\mu\Gamma^{\mu\nu\lambda}\wt{D}_\nu\psi_\lambda.
\eeq
In this expression all the $O(\psi^3)$ terms are hidden in $\wt{D}_\mu$.
To show that the explicit $O(\psi^3)$ terms are canceled
we need Fierz identity \eqref{fierz4}.
Then the equation of motion of $\psi_\mu$ is
\bea
0 & = & \Gamma^{\mu\nu\lambda}\wt{D}_\nu\psi_\lambda
\nn & = & \Gamma^{\mu\nu\lambda}\Big[D_\nu(\hat{\omega})\psi_\lambda
 +\frac{1}{288}(\Gamma^{\mu_1\dots\mu_4}{}_\nu
 -8\delta_\nu{}^{\mu_1}\Gamma^{\mu_2\mu_3\mu_4})\psi_\lambda\hat{F}_{\mu_1\dots\mu_4}
\Big].
\eea

Commutators of the local symmetries
except the one between two supersymmetries are given as follows:
\beq
[\delta^A_{\theta_1},\delta^A_{\theta_2}]=0,\quad
[\delta^A_\theta,\delta^L_\lambda]=0,\quad
[\delta^A_\theta,\delta^S_\xi]=0,
\label{commu1}
\eeq
\begin{align}
[\delta^A_\theta,\delta^D_\epsilon] & =\delta^A_{\theta'}, &
\theta'_{\mu\nu} & =-3\epsilon^\lambda\p_{[\lambda}\theta_{\mu\nu]},
\label{commu2} \\
[\delta^L_{\lambda_1},\delta^L_{\lambda_2}] & =\delta^L_{\lambda_{12}}, &
\lambda_{12}^{ab} & =-[\lambda_1,\lambda_2]^{ab}
 =-\lambda_1{}^a{}_c\lambda_2{}^{cb}+\lambda_2{}^a{}_c\lambda_1{}^{cb},
\label{commu3} \\
[\delta^L_\lambda,\delta^D_\epsilon] & =\delta^L_{\lambda'}, & 
\lambda'_{ab} & =-\epsilon^\mu\p_\mu\lambda_{ab},
\label{commu4} \\
[\delta^L_\lambda,\delta^S_\xi] & =\delta^S_{\xi'}, & 
\xi' & =-\frac{1}{4}\lambda_{ab}\Gamma^{ab}\xi,
\label{commu5} \\
[\delta^D_{\epsilon_1},\delta^D_{\epsilon_2}] & =\delta^D_{\epsilon_{12}}, & 
\epsilon_{12}^\mu & =[\epsilon_1,\epsilon_2]^\mu
 =\epsilon_1{}^\nu\p_\nu\epsilon_2{}^\mu-\epsilon_2{}^\nu\p_\nu\epsilon_1{}^\mu,
\label{commu6} \\
[\delta^D_\epsilon,\delta^S_\xi] & =\delta^S_{\xi''}, & 
\xi'' & =\epsilon^\mu\p_\mu\xi.
\label{commu7}
\end{align}
We see that the above commutators are closed i.e. they are expressed by linear
combinations of the four local symmetry transformations. 
However the commutator between two supersymmetries is not closed:
\beq
[\delta^S_{\xi_1},\delta^S_{\xi_2}] = \delta^D_\epsilon+\delta^L_\lambda
 +\delta^S_\xi+\delta^A_\theta+\delta^t,
\label{commuSS}
\eeq
where
\bea
\epsilon^\mu & = & \frac{i}{4}\bar{\xi}_1\Gamma^\mu\xi_2,
\\
\lambda_{ab} & = & -\epsilon^\mu\hat{\omega}_{\mu ab}
 -\frac{i}{576}\bar{\xi}_1(\Gamma_{ab}{}^{\mu_1\dots\mu_4}\hat{F}_{\mu_1\dots\mu_4}
 +24\Gamma^{\mu\nu}\hat{F}_{ab\mu\nu})\xi_2,
\\
\xi & = & \epsilon^\mu\psi_\mu,
\\
\theta_{\mu\nu} & = & \epsilon^\lambda A_{\lambda\mu\nu}+\frac{i}{4}\bar{\xi}_1\Gamma_{\mu\nu}\xi_2,
\eea
and the `trivial symmetry' $\delta^t$, which is 
proportional to the equation of motion of $\psi_\mu$, is given by
\beq
\delta^t e_\mu{}^a=0,\quad \delta^t A_{\mu\nu\lambda}=0,
\label{trivialea}
\eeq
and
\bea
\delta^t\psi_\mu & = & \frac{i}{16}\bar{\xi}_1\Gamma^\nu\xi_2\Big[
-\frac{5}{12}g_{\nu\mu}\Gamma_\lambda
-\frac{5}{12}g_{\nu\lambda}\Gamma_\mu
+\frac{3}{2}g_{\mu\lambda}\Gamma_\nu
+\frac{29}{144}\Gamma_\mu\Gamma_\nu\Gamma_\lambda
\Big]\Gamma^{\lambda\lambda_1\lambda_2}\wt{D}_{\lambda_1}\psi_{\lambda_2}
\nn & & 
+\frac{i}{32}\bar{\xi}_1\Gamma^{\nu_1\nu_2}\xi_2\Big[
\frac{7}{2}g_{\mu\nu_1}g_{\lambda\nu_2}
-\frac{1}{4}g_{\mu\lambda}\Gamma_{\nu_1\nu_2}
\nn & &
+\frac{1}{3}g_{\lambda\nu_1}\Gamma_\mu\Gamma_{\nu_2}
-\frac{1}{3}g_{\mu\nu_1}\Gamma_{\nu_2}\Gamma_\lambda
+\frac{7}{144}\Gamma_\mu\Gamma_{\nu_1\nu_2}\Gamma_\lambda
\Big]\Gamma^{\lambda\lambda_1\lambda_2}\wt{D}_{\lambda_1}\psi_{\lambda_2}
\nn & &
+\frac{i}{384}\bar{\xi}_1\Gamma^{\nu_1\dots\nu_5}\xi_2\Big[
-g_{\mu\nu_1}g_{\lambda\nu_2}\Gamma_{\nu_3\nu_4\nu_5}
-\frac{1}{12}g_{\mu\nu_1}\Gamma_{\nu_2\dots\nu_5}\Gamma_\lambda
-\frac{1}{12}g_{\lambda\nu_1}\Gamma_\mu\Gamma_{\nu_2\dots\nu_5}
\nn & &
+\frac{1}{144}\Gamma_\mu\Gamma_{\nu_1\dots\nu_5}\Gamma_\lambda
\Big]\Gamma^{\lambda\lambda_1\lambda_2}\wt{D}_{\lambda_1}\psi_{\lambda_2}.
\label{trivialpsi}
\eea
To show \eqref{commuSS} on $A_{\mu\nu\lambda}$ we need Fierz identity \eqref{fierz1}.

Thus the symmetry algebra is open. Moreover the 3-form gauge transformation
is reducible i.e. the transformation parameter also has a `local symmetry'
$\delta\theta_{\mu\nu}=2\p_{[\mu}\theta_{\nu]}$. Again $\theta_\mu$ also has a
`local symmetry' $\delta\theta_\mu=\p_\mu\theta$.
Therefore to perform gauge fixing of those local symmetries 
we must use field-antifield formalism. 

\section{Ghosts and commutators}

In this section we introduce ghosts into $D=11$ supergravity 
following the field-antifield formalism (mainly following the description in \cite{gps94}),
and investigate Jacobi identity of the local symmetry algebra.

$e_\mu{}^a, \psi_\mu^\alpha$, and $A_{\mu\nu\lambda}$
in the original supergravity action $S_0$ are denoted collectively
by $C^{A_{-1}}$, where indices $A_{-1}, B_{-1},\dots$ denote 
three types of fields $(e), (\psi)$, and $(A)$:
\bea
(C^{A_{-1}}) & = & (C^{(e)},C^{(\psi)},C^{(A)}) \nn
 & = & (C^{(\mu a)},C^{(\mu\alpha)},C^{[\mu\nu\lambda]})
 =(e_\mu{}^a, \psi_\mu^\alpha, A_{\mu\nu\lambda}).
\eea
In our notation, indices implicitly contain spacetime positions, and
contractions of such indices contain integrations of the positions.
Usually we do not have to be conscious of the presence of these integrations and
we can think of indices as those taking discrete values. 
However when derivative operators are involved we have to deal with them carefully,
as is done in Appendix B.
In this notation, the equation of motion of $\psi_\mu$ is 
\beq
0=\p{\cal S}_0/\p\psi_\mu^\alpha
 =-ie({\cal C}^{-1}\Gamma^{\mu\nu\lambda}\wt{D}_\nu\psi_\lambda)_\alpha.
\eeq
The infinitesimal local symmetry transformation with parameter $\epsilon^{A_0}$ is denoted by
\beq
\delta_\epsilon C^{A_{-1}}=R^{A_{-1}}{}_{A_0}[C^{B_{-1}}]\epsilon^{A_0}.
\eeq
$R^{A_{-1}}{}_{A_0}$ may contain derivative operators, and depend on 
$C^{B_{-1}}$. Explicit expressions of $R^{A_{-1}}{}_{A_0}$ are
readily read off from \eqref{symse}-\eqref{syma}.
Indices $A_0, B_0, \dots$ denote four types of symmetries $(A), (L), (D)$ and $(S)$:
\bea
(\epsilon^{A_0}) & = & (\epsilon^{(A)},\epsilon^{(L)},\epsilon^{(D)},\epsilon^{(S)}) \nn
 & = & (\epsilon^{[\mu\nu]},\epsilon^{[ab]},\epsilon^{(\mu)},\epsilon^{(\alpha)})
 =(\theta_{\mu\nu}, \lambda^{ab}, \epsilon^\mu, \xi^\alpha).
\eea
The action ${\cal S}_0$ is invariant under the transformation:
\beq
\p{\cal S}_0/\p C^{A_{-1}}R^{A_{-1}}{}_{A_0}=0.
\eeq
Corresponding to the symmetry, we introduce ghosts $C^{A_0}$:
\bea
(C^{A_0}) & = & (C^{[\mu\nu]},C^{[ab]},C^{(\mu)},C^{(\alpha)})
\nn & = & (c_{\mu\nu},c^{ab},c^\mu,c^\alpha).
\eea
To avoid confusion, $e_{\mu a}e_{\nu b}c^{ab}$ will never be denoted by $c_{\mu\nu}$.
Since the local symmetry is reducible i.e. the symmetry parameter $\theta_{\mu\nu}$ has a `symmetry',
there exist such $R^{A_0}{}_{A_1}$ and $R^{A_1}{}_{A_2}$ that
\beq
R^{A_{-1}}{}_{A_0}R^{A_0}{}_{A_1}=0,\quad R^{A_0}{}_{A_1}R^{A_1}{}_{A_2}=0.
\eeq
Correspondingly we introduce a `ghost of ghost' $C^{A_1}$
and a `ghost of ghost of ghost' $C^{A_2}$. 
Indices $A_1$ and $A_2$ take only one type of fields
respectively: $A_1=[\mu]$, and $A_2$ is empty:
\beq
(C^{A_1})=(C^{[\mu]})=(C_{\mu}), \quad (C^{A_2})=(C).
\eeq
$R^{A_0}{}_{A_1}$ is nonzero only when $A_0=(A)=[\mu\nu]$.
For $A_0=[\mu\nu]$ and $A_1=[\mu]$, explicit expressions of
$R^{A_0}{}_{A_1}$ and $R^{A_1}{}_{A_2}$ are given by
\beq
R^{A_0}{}_{B_1}C^{B_1}=2\p_{[\mu}C_{\nu]}, \quad
R^{A_1}{}_{B_2}C^{B_2}=\p_{\mu}C,
\eeq
and note that $R^{A_0}{}_{A_1}$ and $R^{A_1}{}_{A_2}$ do not depend on fields.

We assign statistical parity st and ghost number gh to each field as follows: 
\beq
\text{st}[C^{A_n}]=A_n+n+1,\quad
\text{gh}[C^{A_n}]=n+1.
\eeq
As usual,
st$[f]=0 (=1)$ mod 2 means that $f$ is commuting (anticommuting).
Statistical parity of an index $A_n$ is denoted by $A_n$ itself.
If $A_n$ contains a spinor index, then $A_n=1$, otherwise $A_n=0$.
We will often use this notation in sign factors, especially in powers of $(-1)$.

The commutator of two local symmetries is
\bea
[\delta_{\epsilon_1}, \delta_{\epsilon_2}]C^{A_{-1}}
 & = & \Big[\p R^{A_{-1}}{}_{B_0}/\p C^{B_{-1}}R^{B_{-1}}{}_{C_0}
\nn & &
  -(-1)^{B_0C_0}\p R^{A_{-1}}{}_{C_0}/\p C^{B_{-1}}R^{B_{-1}}{}_{B_0}\Big]
 \epsilon_1^{C_0}\epsilon_2^{B_0},
\label{commu0}
\eea
and the right hand side must be expressed by linear combination of 
the local symmetries and `trivial symmetry' proportional to the equation of motion
$\p{\cal S}_0/\p C^{A_{-1}}$:
\bea
\p R^{A_{-1}}{}_{B_0}/\p C^{B_{-1}}R^{B_{-1}}{}_{C_0}
  -(-1)^{B_0C_0}\p R^{A_{-1}}{}_{C_0}/\p C^{B_{-1}}R^{B_{-1}}{}_{B_0}
\nn 
=R^{A_{-1}}{}_{A_0}T^{A_0}{}_{B_0C_0}+\p{\cal S}_0/\p C^{B_{-1}}E^{B_{-1}A_{-1}}{}_{B_0C_0},
\label{defTE}
\eea
where $T^{A_0}{}_{B_0C_0}$ and $E^{B_{-1}A_{-1}}{}_{B_0C_0}$ has 
graded antisymmetry in $(B_0,C_0)$ and $(B_{-1},A_{-1})$:
\bea
T^{A_0}{}_{B_0C_0} & = & (-1)^{1+B_0C_0}T^{A_0}{}_{C_0B_0}, \\
E^{B_{-1}A_{-1}}{}_{B_0C_0} & = & (-1)^{1+A_{-1}B_{-1}}E^{A_{-1}B_{-1}}{}_{B_0C_0} \nn
 & = & (-1)^{1+B_0C_0}E^{B_{-1}A_{-1}}{}_{C_0B_0}.
\eea
$T^{A_0}{}_{B_0C_0}$ is the `structure constant' of this symmetry, and
its definition has an ambiguity: If we add $R^{A_0}{}_{A_1}\wt{T}^{A_1}{}_{B_0C_0}$
to $T^{A_0}{}_{B_0C_0}$, \eqref{defTE} is still satisfied.
An explicit form of $T^{A_0}{}_{B_0C_0}$ up to this ambiguity can be read off from 
\eqref{commu1}-\eqref{commuSS}. Defining $T^{A_0}$ as
\beq
T^{A_0}:=(-1)^{B_0}T^{A_0}{}_{B_0C_0}C^{C_0}C^{B_0},
\eeq
and $\bar{c}_\alpha:=-(c^T{\cal C}^{-1})_\alpha$, 
components of $T^{A_0}$ are
\bea
T^{[\mu\nu]} & = & 6c^\lambda\p_{[\lambda}c_{\mu\nu]}
 -\frac{i}{4}\bar{c}\Gamma^\lambda c A_{\lambda\mu\nu}
 -\frac{i}{4}\bar{c}\Gamma_{\mu\nu}c,
\label{defTA}
\\
T^{[ab]} & = & -2c^a{}_cc^{cb}+2c^\mu\p_\mu c^{ab}
 +\frac{i}{4}\bar{c}\Gamma^\mu c \hat{\omega}_\mu{}^{ab}
\nn & &
 +\frac{i}{576}\bar{c}(\Gamma^{ab\mu_1\dots\mu_4}
 +24e^{a\mu_1}e^{b\mu_2}\Gamma^{\mu_3\mu_4})c \hat{F}_{\mu_1\dots\mu_4},
\label{defTL}
\\
T^{(\mu)} & = & 2c^\nu\p_\nu c^\mu-\frac{i}{4}\bar{c}\Gamma^\mu c,
\label{defTD}
\\
T^{(\alpha)} & = & \frac{1}{2}c_{ab}(\Gamma^{ab}c)^\alpha-2c^\mu\p_\mu c^\alpha
 -\frac{i}{4}\bar{c}\Gamma^\mu c\psi_\mu^\alpha,
\label{defTS}
\eea
where the ambiguity is fixed so that $T^{A_0}$ contains $c_{\mu\nu}$ only in the form of its field strength.
Similarly $E^{B_{-1}A_{-1}}{}_{B_0C_0}$ can be read off from \eqref{trivialea} and \eqref{trivialpsi}.
$E^{B_{-1}A_{-1}}{}_{B_0C_0}$ also has an ambiguity: if we add 
$\p{\cal S}_0/\p C^{C_{-1}}\wt{E}^{C_{-1}B_{-1}A_{-1}}{}_{B_0C_0}$ with 
$(C_{-1},B_{-1})$ gradedly antisymmetrized to $E^{B_{-1}A_{-1}}{}_{B_0C_0}$,
\eqref{defTE} is still satisfied. For simplicity we fix this kind of ambiguity so that no
terms proportional to derivatives of fields appear. Then
defining $E^{B_{-1}A_{-1}}$ as
\beq
E^{B_{-1}A_{-1}}:=(-1)^{B_0}E^{B_{-1}A_{-1}}{}_{B_0C_0}C^{C_0}C^{B_0},
\eeq
explicit forms of nonzero components of $E^{B_{-1}A_{-1}}$ are given by
\bea
E^{(\nu\beta)(\mu\alpha)} & = & \frac{1}{16}e^{-1}\Big[\bar{c}\Gamma^a c\Big\{
-\frac{5}{12}e_{\mu a}\Gamma_\nu {\cal C}
-\frac{5}{12}e_{\nu a}\Gamma_\mu {\cal C}
+\frac{3}{2}e_{\mu\nu}\Gamma_a {\cal C}
+\frac{29}{144}\Gamma_\mu\Gamma_a\Gamma_\nu {\cal C}
\Big\}^{\alpha\beta}
\nn & & 
+\frac{1}{2}\bar{c}\Gamma^{a_1a_2}c\Big\{
\frac{7}{2}e_{\mu a_1}e_{\nu a_2}{\cal C}
-\frac{1}{4}g_{\mu\nu}\Gamma_{a_1a_2}{\cal C}
\nn & &
+\frac{1}{3}e_{\nu a_1}\Gamma_\mu\Gamma_{a_2}{\cal C}
-\frac{1}{3}e_{\mu a_1}\Gamma_{a_2}\Gamma_\nu {\cal C}
+\frac{7}{144}\Gamma_\mu\Gamma_{a_1a_2}\Gamma_\nu {\cal C}
\Big\}^{\alpha\beta}
\nn & &
+\frac{1}{24}\bar{c}\Gamma^{a_1\dots a_5}c\Big\{
-e_{\mu a_1}e_{\nu a_2}\Gamma_{a_3a_4a_5}{\cal C}
\nn & &
-\frac{1}{12}e_{\mu a_1}\Gamma_{a_2\dots a_5}\Gamma_\nu {\cal C}
-\frac{1}{12}e_{\nu a_1}\Gamma_\mu\Gamma_{a_2\dots a_5}{\cal C}
+\frac{1}{144}\Gamma_\mu\Gamma_{a_1\dots a_5}\Gamma_\nu {\cal C}
\Big\}^{\alpha\beta}
\Big].
\label{defE}
\eea
Note that $E^{B_{-1}A_{-1}}{}_{B_0C_0}$ has the following properties, 
which will often be used later:
\bea
E^{(\psi)(\psi)}{}_{(S)(S)}~\text{is the only nonzero component} \nn
 \text{and it depends only on $e_\mu{}^a$}.
\label{Econd}
\eea

Next let us investigate Jacobi identity
\beq
([\delta_{\epsilon_1}, [\delta_{\epsilon_2}, \delta_{\epsilon_3}]]
+[\delta_{\epsilon_2}, [\delta_{\epsilon_3}, \delta_{\epsilon_1}]]
+[\delta_{\epsilon_3}, [\delta_{\epsilon_1}, \delta_{\epsilon_2}]])
C^{A_{-1}}=0.
\eeq
The left hand side can be calculated using \eqref{commu0} and \eqref{defTE}, and we obtain
\beq
0=R^{A_{-1}}{}_{A_0}A^{A_0}{}_{B_0C_0D_0}
 +\p{\cal S}_0/\p C^{B_{-1}}B^{B_{-1}A_{-1}}{}_{B_0C_0D_0},
\label{jacobi2}
\eeq
where
\bea
A^{A_0}{}_{B_0C_0D_0} & = & \p T^{A_0}{}_{[B_0C_0}/\p C^{A_{-1}}R^{A_{-1}}{}_{D_0\}}
 -T^{A_0}{}_{[B_0|E_0|}T^{E_0}{}_{C_0D_0\}},
\label{defA}
\eea
\bea
B^{B_{-1}A_{-1}}{}_{B_0C_0D_0} & = & 
 \p E^{B_{-1}A_{-1}}{}_{[B_0C_0}/\p C^{D_{-1}} R^{D_{-1}}{}_{D_0\}}
 -E^{B_{-1}A_{-1}}{}_{[B_0|E_0|}T^{E_0}{}_{C_0D_0\}}
\nn & &
 -(-1)^{A_{-1}B_0}\p R^{B_{-1}}{}_{[B_0}/\p C^{D_{-1}}E^{D_{-1}A_{-1}}{}_{C_0D_0\}}
\nn & &
 +(-1)^{A_{-1}B_{-1}+B_{-1}B_0}\p R^{A_{-1}}{}_{[B_0}/\p C^{D_{-1}}E^{D_{-1}B_{-1}}{}_{C_0D_0\}},
\label{defB}
\eea
and $[B_0C_0D_0\}$ means graded antisymmetrization. 
It is understood that if there is an sign factor dependent on these indices,
we also interchange the indices in the sign factor. For example,
\begin{multline}
(-1)^{A_{-1}B_0}\p R^{B_{-1}}{}_{[B_0}/\p C^{D_{-1}}E^{D_{-1}A_{-1}}{}_{C_0D_0\}}
\\
= 
 \frac{1}{3}\Big[(-1)^{A_{-1}B_0}\p R^{B_{-1}}{}_{B_0}/\p C^{D_{-1}}E^{D_{-1}A_{-1}}{}_{C_0D_0}
\\
 +(-1)^{A_{-1}C_0+B_0(C_0+D_0)}\p R^{B_{-1}}{}_{C_0}/\p C^{D_{-1}}E^{D_{-1}A_{-1}}{}_{D_0B_0}
\\
 +(-1)^{A_{-1}D_0+D_0(B_0+C_0)}\p R^{B_{-1}}{}_{D_0}/\p C^{D_{-1}}E^{D_{-1}A_{-1}}{}_{B_0C_0}
\Big].
\end{multline}
To satisfy \eqref{jacobi2}, $A^{A_0}{}_{B_0C_0D_0}$ and $B^{B_{-1}A_{-1}}{}_{B_0C_0D_0}$
must be in the following form (see e.g. \cite{gps94}):
\bea
A^{A_0}{}_{B_0C_0D_0} & = &  R^{A_0}{}_{A_1}F^{A_1}{}_{B_0C_0D_0}
 +\p S_0/\p C^{B_{-1}}D^{B_{-1}A_0}{}_{B_0C_0D_0},
\label{reducedA} \\
B^{B_{-1}A_{-1}}{}_{B_0C_0D_0} & = & 
 (-1)^{A_{-1}A_0}R^{B_{-1}}{}_{A_0}D^{A_{-1}A_0}{}_{B_0C_0D_0}
\nn & &
+(-1)^{1+B_{-1}(A_{-1}+A_0)}R^{A_{-1}}{}_{A_0}D^{B_{-1}A_0}{}_{B_0C_0D_0}
\nn & &
+\p S_0/\p C^{C_{-1}}M^{C_{-1}B_{-1}A_{-1}}{}_{B_0C_0D_0},
\label{reducedB}
\eea
where $F^{A_1}{}_{B_0C_0D_0}$, $D^{A_{-1}A_0}{}_{B_0C_0D_0}$ and $M^{C_{-1}B_{-1}A_{-1}}{}_{B_0C_0D_0}$
has graded antisymmetry in $(B_0,C_0,D_0)$ and $(C_{-1},B_{-1},A_{-1})$.
The definition of $F^{A_1}{}_{B_0C_0D_0}$ has an ambiguity:
If we add $R^{A_1}{}_{A_2}\wt{F}^{A_2}{}_{B_0C_0D_0}$ to $F^{A_1}{}_{B_0C_0D_0}$,
\eqref{reducedA} is still satisfied. $D^{B_{-1}A_0}{}_{B_0C_0D_0}$ and 
$M^{C_{-1}B_{-1}A_{-1}}{}_{B_0C_0D_0}$ also has an ambiguity similar to 
$E^{B_{-1}A_{-1}}{}_{B_0C_0}$, which will be fixed similarly.

By computing the expression \eqref{defA} and \eqref{defB} explicitly, we can confirm
that $A^{A_0}{}_{B_0C_0D_0}$ and $B^{B_{-1}A_{-1}}{}_{B_0C_0D_0}$ are indeed 
in the form of \eqref{reducedA} and \eqref{reducedB}, and obtain explicit expressions of
$F^{A_1}{}_{B_0C_0D_0}$, $D^{A_{-1}A_0}{}_{B_0C_0D_0}$ and $M^{C_{-1}B_{-1}A_{-1}}{}_{B_0C_0D_0}$.
Details of the calculation of $A^{A_0}$ and $B^{B_{-1}A_{-1}}$ are given in Appendix \ref{appb}, and
we find 
\beq
M^{C_{-1}B_{-1}A_{-1}}{}_{B_0C_0D_0}=0.
\eeq
and the following properties:
\bea
F^{A_1}~\text{does not depend on $\psi_\mu^\alpha$ and $c_{ab}$.}
\label{Fcond}
\eea
\bea
D^{(\psi)(L)}{}_{(S)(S)(S)}~\text{is the only nonzero component} \nn
 \text{and it depends only on $e_\mu{}^a$.}
\label{Dcond}
\eea
See \eqref{defF}, and \eqref{defD} or \eqref{redD} for explicit expressions.
These will be used in the next section.

\section{Constructing a master action}

In this section we construct a master action $S$ satisfying the classical master equation.
Following the general theory of field-antifield formalism,
\beq
(C^A)=(C^{A_{-1}}, C^{A_0}, C^{A_1}, C^{A_2}),
\eeq
are called fields,
and we introduce corresponding antifields 
\beq
(C^*_A)=(C^*_{A_{-1}}, C^*_{A_0}, C^*_{A_1}, C^*_{A_2}).
\eeq
Statistical parity st, ghost number gh, and antighost number ag of these fields are
\begin{align}
\text{st}[C^{A_n}] & =A_n+n+1, & \text{st}[C^*_{A_n}] & =A_n+n, \\
\text{gh}[C^{A_n}] & =n+1, & \text{gh}[C^*_{A_n}] & =-n-2, \\
\text{ag}[C^{A_n}] & =0, & \text{ag}[C^*_{A_n}] & =n+2.
\end{align}
The importance of antighost number in order-by-order analysis of master actions
has been pointed out in \cite{fh90}.
The antibracket $(X,Y)$ is defined as
\beq
(X,Y):=\p X/\p C^A\cdot(\p/\p C^*_A)Y - \p X/\p C^*_A\cdot(\p/\p C^A)Y,
\eeq
If $X$ is Grassmann even i.e. st$[X]=0$ mod 2, its `self-antibracket' is
\beq
\frac{1}{2}(X,X)=\p X/\p C^A\cdot(\p/\p C^*_A)X.
\eeq
Starting from the original action $S_0=\frac{1}{2\kappa^2}{\cal S}_0[C^{A_{-1}}]$,
we add new terms which contain antifields, and 
the total action ${\cal S}={\cal S}_0+\dots$ must satisfy the classical master equation:
\beq
({\cal S},{\cal S})=0.
\eeq
To obtain a proper solution to this equation, terms consisting of one antifield and 
one ghost must be given by
\beq
{\cal S}_1=C^*_{A_{-1}}R^{A_{-1}}{}_{A_0}C^{A_0}
 +C^*_{A_0}R^{A_0}{}_{A_1}C^{A_1}
 +C^*_{A_1}R^{A_1}{}_{A_2}C^{A_2}.
\label{sol1}
\eeq
To see what terms we should add next, let us compute the
self-antibracket of ${\cal S}_0+{\cal S}_1$:
\bea
\frac{1}{2}({\cal S}_0+{\cal S}_1,{\cal S}_0+{\cal S}_1) & = & 
 (-1)^{1+B_0}C^*_{B_{-1}}\p R^{B_{-1}}{}_{B_0}/\p C^{A_{-1}}
  R^{A_{-1}}{}_{A_0}C^{A_0}C^{B_0}
\nn & = & 
 \frac{1}{2}(-1)^{1+B_0}C^*_{A_{-1}}\big[R^{A_{-1}}{}_{A_0}T^{A_0}{}_{B_0C_0}
\nn & &
 +\p S_0/\p C^{B_{-1}}E^{B_{-1}A_{-1}}{}_{B_0C_0}
 \big]C^{C_0}C^{B_0},
\label{action1}
\eea
where we used \eqref{defTE}.
Our procedure to find new terms to be added is the following:
Suppose we have the following expression in the result of the computation of the self-antibracket:
\beq
C^*_{C_{n_1}}C^*_{D_{n_2}}\dots\times C^*_{A_n}R^{A_n}{}_{B_{n+1}}
K^{B_{n+1}C_{n_1}D_{n_2}\dots}{}_{C'_{m_1}D'_{m_2}\dots}[C^{A_{-1}}]
\times C^{C'_{m_1}}C^{D'_{m_2}}\dots,
\label{term1tobecanceled}
\eeq
then we add 
\begin{multline}
(-1)^{1+(C_{n_1}+n_1)+(D_{n_2}+n_2)+\dots}
C^*_{C_{n_1}}C^*_{D_{n_2}}\dots
\\
\times C^*_{A_{n+1}}
K^{A_{n+1}C_{n_1}D_{n_2}\dots}{}_{C'_{m_1}D'_{m_2}\dots}[C^{A_{-1}}]
\times C^{C'_{m_1}}C^{D'_{m_2}}\dots,
\end{multline}
to the action. When indices have graded symmetry we need to
put a combinatorial factor to the above.
If we have the following expression
\beq
C^*_{C_{n_1}}C^*_{D_{n_2}}\dots
\p S_0/\p C^{A_{-1}}
K^{A_{-1}C_{n_1}D_{n_2}\dots}{}_{C'_{m_1}D'_{m_2}\dots}[C^{A_{-1}}]
\times C^{C'_{m_1}}C^{D'_{m_2}}\dots,
\label{term2tobecanceled}
\eeq
then we add 
\begin{multline}
(-1)^{1+(C_{n_1}+n_1)+(D_{n_2}+n_2)+\dots}
C^*_{C_{n_1}}C^*_{D_{n_2}}\dots
\\
\times C^*_{A_{-1}}
K^{A_{-1}C_{n_1}D_{n_2}\dots}{}_{C'_{m_1}D'_{m_2}\dots}[C^{A_{-1}}]
\times C^{C'_{m_1}}C^{D'_{m_2}}\dots,
\end{multline}
to the action. Contributions to the self-antibracket from these new terms cancel
\eqref{term1tobecanceled} and \eqref{term2tobecanceled},
and generate additional contribution.
If it contains terms in the form of \eqref{term1tobecanceled} and \eqref{term2tobecanceled} again
then we can repeat this procedure. As can be seen below,
this procedure generates antighost number expansion of the action.

Let us apply this procedure to \eqref{action1}, which has antighost number 1.
We obtain the following new terms of antighost number 2:
\bea
{\cal S}_{(0;0,0)} & = & \frac{1}{2}(-1)^{B_0}C^*_{A_0}T^{A_0}{}_{B_0C_0}C^{C_0}C^{B_0},
\label{sol2} \\
{\cal S}_{(-1,-1;0,0)} & = & \frac{1}{4}(-1)^{1+A_{-1}+B_0}C^*_{A_{-1}}C^*_{B_{-1}}
 E^{B_{-1}A_{-1}}{}_{B_0C_0}C^{C_0}C^{B_0},
\label{sol3}
\eea
and the self-antibracket of ${\cal S}_{(2)}={\cal S}_0+{\cal S}_1+{\cal S}_{(0;0,0)}+{\cal S}_{(-1,-1;0,0)}$
is
\bea
\frac{1}{2}({\cal S}_{(2)},{\cal S}_{(2)}) & = &
 \frac{1}{2}(-1)^{C_0}C^*_{A_0}A^{A_0}{}_{B_0C_0D_0}
 C^{D_0}C^{C_0}C^{B_0}
\nn & &
 -\frac{1}{4}(-1)^{A_{-1}+C_0}C^*_{A_{-1}}C^*_{B_{-1}}B^{B_{-1}A_{-1}}{}_{B_0C_0D_0}
 C^{D_0}C^{C_0}C^{B_0}
\nn & &
 -\frac{1}{4}(-1)^{B_0+D_0+B_{-1}(B_0+C_0+A_{-1})}C^*_{B_{-1}}C^*_{A_0}
\nn & &
 \times 
 \p T^{A_0}{}_{B_0C_0}/\p C^{A_{-1}}E^{B_{-1}A_{-1}}{}_{D_0E_0}
 C^{E_0}C^{D_0}C^{C_0}C^{B_0}
\nn & &
 -C^*_{A_0}T^{A_0}{}_{B_0C_0}R^{C_0}{}_{A_1}C^{A_1}C^{B_0},
\eea
where we dropped terms which we can easily see vanish from \eqref{Econd}.
We also see that the last term in the above vanishes because $R^{C_0}{}_{A_1}C^{A_1}$ is the gauge transformation of
$C^{C_0=[\mu\nu]}$, and $T^{A_0}{}_{B_0C_0}$ contains $C^{[\mu\nu]}$ only
in the form of its field strength.
Then using \eqref{reducedA} and \eqref{reducedB},
\bea
\frac{1}{2}({\cal S}_{(2)},{\cal S}_{(2)}) & = & 
 \frac{1}{2}(-1)^{C_0}C^*_{A_0}R^{A_0}{}_{A_1}F^{A_1}{}_{B_0C_0D_0}
 C^{D_0}C^{C_0}C^{B_0}
\nn & &
 +\frac{1}{2}(-1)^{C_0}C^*_{A_0}\p S_0/\p C^{B_{-1}}D^{B_{-1}A_0}{}_{B_0C_0D_0}
 C^{D_0}C^{C_0}C^{B_0}
\nn & &
 -\frac{1}{2}(-1)^{A_{-1}+C_0+A_{-1}A_0}C^*_{A_{-1}}C^*_{B_{-1}}
\nn & &
 \times R^{B_{-1}}{}_{A_0}D^{A_{-1}A_0}{}_{B_0C_0D_0}
 C^{D_0}C^{C_0}C^{B_0}
\nn & & 
 -\frac{1}{4}(-1)^{B_0+D_0+B_{-1}(B_0+C_0+A_{-1})}C^*_{B_{-1}}C^*_{A_0}
\nn & & 
 \times\p T^{A_0}{}_{B_0C_0}/\p C^{A_{-1}}E^{B_{-1}A_{-1}}{}_{D_0E_0}
 C^{E_0}C^{D_0}C^{C_0}C^{B_0}.
\label{action2}
\eea
The last term in the above is of antighost number 3, and the rest are of antighost number 2.
Let us cancel the terms of lower antighost number:
To cancel the first term in the above, we introduce the following term:
\beq
{\cal S}_{(1;0,0,0)} = \frac{1}{2}(-1)^{1+C_0}C^*_{A_1}
 F^{A_1}{}_{B_0C_0D_0}C^{D_0}C^{C_0}C^{B_0}.
\label{sol4}
\eeq
To cancel the second and third term in the above, we introduce
\beq
{\cal S}_{(-1,0;0,0,0)} = \frac{1}{2}(-1)^{1+A_0+C_0}C^*_{A_0}C^*_{B_{-1}}
 D^{B_{-1}A_0}{}_{B_0C_0D_0}C^{D_0}C^{C_0}C^{B_0}.
\label{sol5}
\eeq
The self-antibracket of ${\cal S}_{(3)}={\cal S}_{(2)}+{\cal S}_{(1;0,0,0)}+{\cal S}_{(-1,0;0,0,0)}$
is
\bea
\frac{1}{2}({\cal S}_{(3)},{\cal S}_{(3)}) & = & 
 \frac{1}{4}(-1)^{1+D_0+F_0+B_0(C_0+D_0)}C^*_{B_0}C^*_{A_0}
\nn & & 
 \times \p T^{A_0}{}_{C_0D_0}/\p C^{A_{-1}}D^{A_{-1}B_0}{}_{E_0F_0G_0}
 C^{G_0}C^{F_0}C^{E_0}C^{D_0}C^{C_0}
\nn & &
 +\frac{1}{2}(-1)^{B_0+D_0}C^*_{A_1}Z^{A_1}{}_{B_0C_0D_0E_0}
 C^{E_0}C^{D_0}C^{C_0}C^{B_0}
\nn & &
 +\frac{1}{2}(-1)^{A_0+B_0+D_0}C^*_{A_0}C^*_{B_{-1}}W^{B_{-1}A_0}{}_{B_0C_0D_0E_0}
\nn & &
 \times C^{E_0}C^{D_0}C^{C_0}C^{B_0}
\nn & & 
 +\frac{3}{2}(-1)^{1+C_0}C^*_{A_1}
 F^{A_1}{}_{B_0C_0D_0}R^{D_0}{}_{B_1}C^{B_1}C^{C_0}C^{B_0},
\label{action3}
\eea
where we dropped terms which we can easily see vanish
from \eqref{Econd}, \eqref{Fcond} and \eqref{Dcond}. $Z^{A_1}{}_{B_0C_0D_0E_0}$
and $W^{B_{-1}A_0}{}_{B_0C_0D_0E_0}$ are defined as
\beq
Z^{A_1}{}_{B_0C_0D_0E_0} := 
 \p F^{A_1}{}_{[B_0C_0D_0}/\p C^{A_{-1}}R^{A_{-1}}{}_{E_0\}}
 -\frac{3}{2}F^{A_1}{}_{[B_0C_0|F_0|}T^{F_0}{}_{D_0E_0\}},
\label{defZ}
\eeq
\bea
W^{B_{-1}A_0}{}_{B_0C_0D_0E_0} & := &
 \p D^{B_{-1}A_0}{}_{[B_0C_0D_0}/\p C^{A_{-1}}R^{A_{-1}}{}_{E_0\}}
\nn & &
 -\frac{3}{2}D^{B_{-1}A_0}{}_{[B_0C_0|F_0|}T^{F_0}{}_{D_0E_0\}}
\nn & &
 +(-1)^{B_{-1}(A_0+B_0+F_0)}T^{A_0}{}_{[B_0|F_0|}D^{B_{-1}F_0}{}_{C_0D_0E_0\}}
\nn & &
 +(-1)^{A_0B_0}\p R^{B_{-1}}{}_{[B_0}/\p C^{A_{-1}}D^{A_{-1}A_0}{}_{C_0D_0E_0\}}
\nn & &
 +\frac{1}{2}(-1)^{1+B_{-1}(A_0+B_0+C_0+A_{-1})}
  \p T^{A_0}{}_{[B_0C_0}/\p C^{A_{-1}}E^{B_{-1}A_{-1}}{}_{D_0E_0\}}.
\label{defW}
\eea
The last term in \eqref{action3} vanishes, because
$R^{D_0}{}_{B_1}C^{B_1}$ is the gauge transformation of $C^{D_0=[\mu\nu]}$,
and $F^{A_1}{}_{B_0C_0D_0}$ contains $C^{[\mu\nu]}$ only in the form of its field strength.

In Appendix \ref{appc}, we show that
\bea
Z^{A_1}{}_{B_0C_0D_0E_0} & = & R^{A_1}{}_{A_2}Y^{A_2}{}_{B_0C_0D_0E_0},
\\
W^{B_{-1}A_0}{}_{C_0D_0E_0F_0} & = & R^{B_{-1}}{}_{B_0}V^{B_0A_0}{}_{C_0D_0E_0F_0},
\eea
and explicit expressions of $Y^{A_2}{}_{C_0D_0E_0F_0}$ and $V^{B_0A_0}{}_{C_0D_0E_0F_0}$ are
given by \eqref{defY}, and \eqref{defV} or \eqref{redV}.
$Y^{A_2}{}_{C_0D_0E_0F_0}$ and $V^{B_0A_0}{}_{C_0D_0E_0F_0}$ have graded antisymmetry
in $(C_0,D_0,E_0,F_0)$, and graded symmetry in $(A_0,B_0)$.
$Y^{A_2}$ and $V^{B_0A_0}$ are defined as
\bea
Y^{A_2} & := &
 (-1)^{B_0+D_0}Y^{A_2}{}_{B_0C_0D_0E_0}C^{E_0}C^{D_0}C^{C_0}C^{B_0},
\\
V^{B_0A_0} & := & 
 (-1)^{C_0+E_0}V^{B_0A_0}{}_{C_0D_0E_0F_0}
 C^{F_0}C^{E_0}C^{D_0}C^{C_0}.
\eea
$Y^{A_2}$ and $V^{B_0A_0}{}_{C_0D_0E_0F_0}$ have the following properties:
\bea
Y^{A_2}~\text{does not depend on $\psi_\mu^\alpha$ and $c_{[ab]}$.}
\label{Ycond}
\eea
\bea
V^{(L)(L)}{}_{(S)(S)(S)(S)}~\text{is the only nonzero component}
\nn \text{and it depends only on $e_\mu{}^a$.}
\label{Vcond}
\eea
Then,
\bea
\frac{1}{2}({\cal S}_{(3)},{\cal S}_{(3)}) & = & 
 \frac{1}{4}(-1)^{1+D_0+F_0+B_0(C_0+D_0)}C^*_{B_0}C^*_{A_0}
\nn & & 
 \times \p T^{A_0}{}_{C_0D_0}/\p C^{A_{-1}}D^{A_{-1}B_0}{}_{E_0F_0G_0}
 C^{G_0}C^{F_0}C^{E_0}C^{D_0}C^{C_0}
\nn & &
 +\frac{1}{2}(-1)^{B_0+D_0}C^*_{A_1}R^{A_1}{}_{A_2}Y^{A_2}{}_{B_0C_0D_0E_0}
 C^{E_0}C^{D_0}C^{C_0}C^{B_0}
\nn & &
 +\frac{1}{2}(-1)^{A_0+C_0+E_0}C^*_{A_0}C^*_{B_{-1}}R^{B_{-1}}{}_{B_0}V^{B_0A_0}{}_{C_0D_0E_0F_0}
\nn & &
 \times C^{F_0}C^{E_0}C^{D_0}C^{C_0}.
\eea
The first term in the above is of antighost number 4, and the rest are of antighost number 3.
The terms of lower antighost number are canceled by introducing
\bea
{\cal S}_{(2;0,0,0,0)} & = & \frac{1}{2}(-1)^{1+B_0+D_0}C^*_{A_2}Y^{A_2}{}_{B_0C_0D_0E_0}
 C^{E_0}C^{D_0}C^{C_0}C^{B_0}, \\
\label{sol6}
{\cal S}_{(0,0;0,0,0,0)} & = & \frac{1}{4}(-1)^{1+C_0+E_0}
 C^*_{A_0}C^*_{B_0}V^{B_0A_0}{}_{C_0D_0E_0F_0}
 C^{F_0}C^{E_0}C^{D_0}C^{C_0},
\label{sol7}
\eea
and the self-antibracket of
${\cal S}_{(4)}={\cal S}_{(3)}+{\cal S}_{(2;0,0,0,0)}+{\cal S}_{(0,0;0,0,0,0)}$ is
\begin{multline}
\frac{1}{2}({\cal S}_{(4)},{\cal S}_{(4)}) = 
 \frac{1}{2}(-1)^{1+C_0+E_0}C^*_{A_2}X^{A_2}{}_{B_0C_0D_0E_0F_0}
 C^{F_0}C^{E_0}C^{D_0}C^{C_0}C^{B_0}
\\
+\frac{1}{4}(-1)^{1+D_0+F_0}C^*_{A_0}C^*_{B_0}U^{B_0A_0}{}_{C_0D_0E_0F_0G_0}
 C^{G_0}C^{F_0}C^{E_0}C^{D_0}C^{C_0},
\end{multline}
where we dropped terms which we can easily see vanish from \eqref{Econd}, \eqref{Fcond},
\eqref{Dcond}, \eqref{Ycond}, and \eqref{Vcond}.
$X^{A_2}{}_{B_0C_0D_0E_0F_0}$ and $U^{B_0A_0}{}_{C_0D_0E_0F_0G_0}$ are defined as
\beq
X^{A_2}{}_{B_0C_0D_0E_0F_0} := 
\p Y^{A_2}{}_{[B_0C_0D_0E_0}/\p C^{A_{-1}}R^{A_{(-1)}}{}_{F_0\}}
 -2Y^{A_2}{}_{[B_0C_0D_0|G_0|}T^{G_0}{}_{E_0F_0\}},
\label{defX}
\eeq
\bea
U^{B_0A_0}{}_{C_0D_0E_0F_0G_0} & := &
\p V^{B_0A_0}{}_{[C_0D_0E_0F_0}/\p C^{A_{-1}}R^{A_{-1}}{}_{G_0\}}
\nn & &
-2V^{B_0A_0}{}_{[C_0D_0E_0|H_0|}T^{H_0}{}_{F_0G_0\}}
\nn & &
-(-1)^{A_0C_0}T^{B_0}{}_{[C_0|H_0|}V^{H_0A_0}{}_{D_0E_0F_0G_0\}}
\nn & &
-(-1)^{A_0B_0+B_0C_0}T^{A_0}{}_{[C_0|H_0|}V^{H_0B_0}{}_{D_0E_0F_0G_0\}}
\nn & &
+\frac{1}{2}(-1)^{A_0(C_0+D_0)}\p T^{B_0}{}_{[C_0D_0}/\p C^{A_{-1}}D^{A_{-1}A_0}{}_{E_0F_0G_0\}}
\nn & &
+\frac{1}{2}(-1)^{A_0B_0+B_0(C_0+D_0)}\p T^{A_0}{}_{[C_0D_0}/\p C^{A_{-1}}D^{A_{-1}B_0}{}_{E_0F_0G_0\}}.
\label{defU}
\eea
In Appendix \ref{appd}, we show $X^{A_2}{}_{B_0C_0D_0E_0F_0}=0$ and
$U^{B_0A_0}{}_{C_0D_0E_0F_0G_0}=0$,
which means that ${\cal S}_{(4)}$ is a solution to the master equation ${\cal S}$:
\beq
{\cal S}={\cal S}_{(4)}.
\eeq

\section{Conclusion}

We have constructed an explicit expression of the master action of $D=11$ supergravity.
For readers' convenience we summarize our result:
A solution to the classical master equation $S$ is 
\bea
2\kappa^2 S & = & {\cal S}_0+{\cal S}_1
 +{\cal S}_{(0;0,0)}+{\cal S}_{(-1,-1;0,0)}
\nn & &
 +{\cal S}_{(1;0,0,0)}+{\cal S}_{(-1,0;0,0,0)}
 +{\cal S}_{(2;0,0,0,0)}+{\cal S}_{(0,0;0,0,0,0)},
\eea
where ${\cal S}_0$, ${\cal S}_1$,
 ${\cal S}_{(0;0,0)}$, ${\cal S}_{(-1,-1;0,0)}$, 
 ${\cal S}_{(1;0,0,0)}$, ${\cal S}_{(-1,0;0,0,0)}$, 
 ${\cal S}_{(2;0,0,0,0)}$, ${\cal S}_{(0,0;0,0,0,0)}$
are given by \eqref{sol0}, \eqref{sol1}, \eqref{sol2}, \eqref{sol3}, \eqref{sol4}, 
\eqref{sol5}, \eqref{sol6}, and \eqref{sol7} respectively,
and symbols used in them are defined by \eqref{defTA}-\eqref{defTS}, \eqref{defE}
\eqref{defF}, \eqref{defD}, \eqref{defY}, \eqref{defV} (or, \eqref{redD} and \eqref{redV}).

Note that the master action has an ambiguity that terms in the form of 
canonical transformation can be added. Up to this ambiguity our solution is essentially unique,
as has been shown in \cite{fh90} generally. 
To construct a gauge fixed action, we have to give a gauge fixing fermion, and
introduce more field-antifield pairs
and more terms to the action, but it is completely straightforward. (See e.g. \cite{gps94}.) 

\vs{.5cm}
\noindent
{\large\bf Acknowledgments}\\[.2cm]
I would like to thank S.\ Sararu for correspondence.

\renewcommand{\theequation}{\Alph{section}.\arabic{equation}}
\appendix
\addcontentsline{toc}{section}{Appendix}
\vs{.5cm}
\noindent
{\Large\bf Appendix}
\section{Notation about spinors and Fierz transformation}
\label{appa}
\setcounter{equation}{0}

In this appendix we summarize our notation about spinors, and explain about details on
Fierz transformation.
$32\times 32$ gamma matrices $(\Gamma^a)^\alpha{}_\beta$ in eleven dimensions satisfy
\beq
\{\Gamma^a,\Gamma^b\}=2\eta^{ab},\quad
\eta^{ab}=\text{diag}(-1,+1,\dots,+1),
\eeq
and the charge conjugation matrix ${\cal C}^{\alpha\beta}$ is an antisymmetric matrix
obeying
${\cal C}^{-1}\Gamma^a{\cal C}=-(\Gamma^a)^T$.
10th gamma matrix $\Gamma^{10}$ is given by the product of other gamma matrices:
\beq
\Gamma^{10}=\sigma\Gamma^0\Gamma^1\dots\Gamma^9,\quad \sigma=\pm 1.
\label{sgng10}
\eeq
A Majorana spinor $\psi$ satisfies the following relation:
\beq
\bar{\psi}=\psi^\dagger\Gamma^0=-\psi^T{\cal C}^{-1}.
\eeq
The totally antisymmetric tensor is defined as
\beq
\epsilon^{a_1\dots a_{11}}=\text{sgn}(a_1, \dots, a_{11}),
\eeq
where $\text{sgn}(a_1, \dots, a_{11})$ is the sign of the permutation
$(a_1, \dots, a_{11})$. Then
\beq
\epsilon^{\mu_1\dots\mu_{11}}=e^{-1}\text{sgn}(\mu_1, \dots, \mu_{11}).
\eeq
Gamma matrices have a `duality' relation:
\beq
\Gamma^{a_1\dots a_n}=\frac{\sigma(-1)^{\frac{1}{2}(n+1)(n+2)}}{(11-n)!}
 \epsilon^{a_1\dots a_{11}}\Gamma_{a_{n+1}\dots a_{11}},
\eeq
which means that $\Gamma_{a_1\dots a_n}~(n=0,1,\dots,5)$
are independent and higher gamma matrices can be expressed by the lower ones.
Therefore we can take ${\cal C}^{-1}\Gamma_{a_1\dots a_n}~(n=0,1,\dots,5)$
as a basis of $32\times 32$ matrices. For $n=1,2$, and 5 these matrices are symmetric,
and others are antisymmetric.

We often need to perform Fierz transformation.
Especially we often need to show that sums of terms in the form of
$\bar{c}\Gamma_1c\cdot\bar{c}\Gamma_2, ~
\bar{c}\Gamma_1c\cdot\bar{c}\Gamma_2c$, and
$\bar{c}\Gamma_1c\cdot\bar{c}\Gamma_2c\cdot\bar{c}\Gamma_3$ cancel,
where $\Gamma_1, \Gamma_2$ and $\Gamma_3$ are products of gamma matrices and $c^\alpha$
is a Grassmann even spinor. 
For this purpose the followings can be used:
\begin{multline}
(\Gamma_1)^\alpha{}_{(\beta}({\cal C}^{-1}\Gamma_2)_{\gamma\delta)}
 = \frac{1}{3}\Big[(\Gamma_1)^\alpha{}_{\beta}({\cal C}^{-1}\Gamma_2)_{\gamma\delta}
\\
 +\sum_{n=1,2,5}\frac{(-1)^{n(n-1)/2}}{16n!}(\Gamma_1\Gamma^{a_1\dots a_n}\Gamma_2)^\alpha{}_\beta
  ({\cal C}^{-1}\Gamma_{a_1\dots a_n})_{\gamma\delta}
\Big],
\label{ft3}
\end{multline}
\begin{multline}
({\cal C}^{-1}\Gamma_1)_{(\alpha\beta}({\cal C}^{-1}\Gamma_2)_{\gamma\delta)}
 = \frac{1}{6}\Big[({\cal C}^{-1}\Gamma_1)_{\alpha\beta}({\cal C}^{-1}\Gamma_2)_{\gamma\delta}
 +({\cal C}^{-1}\Gamma_2)_{\alpha\beta}({\cal C}^{-1}\Gamma_1)_{\gamma\delta}
\\
 +\sum_{n=1,2,5}\frac{(-1)^{n(n-1)/2}}{16n!}\{
 ({\cal C}^{-1}\Gamma_1\Gamma^{a_1\dots a_n}\Gamma_2)_{(\alpha\beta)}
 +({\cal C}^{-1}\Gamma_2\Gamma^{a_1\dots a_n}\Gamma_1)_{(\alpha\beta)}
 \}({\cal C}^{-1}\Gamma_{a_1\dots a_n})_{\gamma\delta}
\Big],
\label{ft4}
\end{multline}
\bea
({\cal C}^{-1}\Gamma_1)_{(\alpha\beta}({\cal C}^{-1}\Gamma_2)_{\gamma\delta}
({\cal C}^{-1}\Gamma_3)_{\epsilon)\zeta}
 = \frac{1}{5!}\Big[
 4({\cal C}^{-1}\Gamma_1)_{\alpha\beta}({\cal C}^{-1}\Gamma_2)_{\gamma\delta}
 ({\cal C}^{-1}\Gamma_3)_{\epsilon\zeta}
& & \nn +\frac{1}{8}\sum_{n=1,2,5}\frac{(-1)^{n(n-1)/2}}{n!}
 ({\cal C}^{-1}\Gamma_1\Gamma_{a_1\dots a_n}\Gamma_2)_{(\alpha\beta)}
 ({\cal C}^{-1}\Gamma^{a_1\dots a_n})_{\gamma\delta}
 ({\cal C}^{-1}\Gamma_3)_{\epsilon\zeta}
& & \nn +\frac{1}{8}\sum_{n=1,2,5}\frac{(-1)^{n(n-1)/2}}{n!}
 ({\cal C}^{-1}\Gamma^{a_1\dots a_n})_{\alpha\beta}
 ({\cal C}^{-1}\Gamma_1\Gamma_{a_1\dots a_n}\Gamma_2)_{(\gamma\delta)}
 ({\cal C}^{-1}\Gamma_3)_{\epsilon\zeta}
& & \nn +\frac{1}{4}\sum_{n=1,2,5}\frac{(-1)^{n(n-1)/2}}{n!}
 ({\cal C}^{-1}\Gamma_1)_{\alpha\beta}
 ({\cal C}^{-1}\Gamma^{a_1\dots a_n})_{\gamma\delta}
 ({\cal C}^{-1}\Gamma_2\Gamma_{a_1\dots a_n}\Gamma_3)_{\epsilon\zeta}
& & \nn +\frac{1}{4}\sum_{n=1,2,5}\frac{(-1)^{n(n-1)/2}}{n!}
 ({\cal C}^{-1}\Gamma^{a_1\dots a_n})_{\alpha\beta}
 ({\cal C}^{-1}\Gamma_1)_{\gamma\delta}
 ({\cal C}^{-1}\Gamma_2\Gamma_{a_1\dots a_n}\Gamma_3)_{\epsilon\zeta}
& & \nn +\frac{1}{64}\sum_{n=1,2,5}\sum_{m=1,2,5}
 \frac{(-1)^{n(n-1)/2}}{n!}\frac{(-1)^{m(m-1)/2}}{m!}
& & \nn \times
 ({\cal C}^{-1}\Gamma_1\Gamma_{b_1\dots b_m}\Gamma_{a_1\dots a_n})_{(\alpha\beta)}
 ({\cal C}^{-1}\Gamma^{b_1\dots b_m})_{\gamma\delta}
 ({\cal C}^{-1}\Gamma_2\Gamma^{a_1\dots a_n}\Gamma_3)_{\epsilon\zeta}
& & \nn +\frac{1}{64}\sum_{n=1,2,5}\sum_{m=1,2,5}
 \frac{(-1)^{n(n-1)/2}}{n!}\frac{(-1)^{m(m-1)/2}}{m!}
& & \nn \times
 ({\cal C}^{-1}\Gamma^{b_1\dots b_m})_{\alpha\beta}
 ({\cal C}^{-1}\Gamma_1\Gamma_{b_1\dots b_m}\Gamma_{a_1\dots a_n})_{(\gamma\delta)}
 ({\cal C}^{-1}\Gamma_2\Gamma^{a_1\dots a_n}\Gamma_3)_{\epsilon\zeta}
& & \nn \Big]+(\Gamma_1\leftrightarrow\Gamma_2).
\label{ft5}
\eea
Applying these transformations we obtain, for example, the following Fierz identities:
\bea
0 & = & (\Gamma^b{\cal C})^{(\alpha\beta}(\Gamma_{ab}{\cal C})^{\gamma\delta)},
\label{fierz1} \\
0 & = & (\Gamma^b{\cal C})^{(\alpha\beta}(\Gamma_{ba_1\dots a_4}{\cal C})^{\gamma\delta)}
 -3(\Gamma_{[a_1a_2}{\cal C})^{(\alpha\beta}(\Gamma_{a_3a_4]}{\cal C})^{\gamma\delta)},
\label{fierz2} \\
0 & = & (\Gamma^{bc}{\cal C})^{\alpha(\beta}(\Gamma_{bca_1\dots a_4}{\cal C})^{\gamma\delta)}
 -2(\Gamma_{a_1\dots a_4b}{\cal C})^{\alpha(\beta}(\Gamma^b{\cal C})^{\gamma\delta)}
\nn & &
 -16(\Gamma_{[a_1a_2a_3}{\cal C})^{\alpha(\beta}(\Gamma_{a_4]}{\cal C})^{\gamma\delta)}
 +24(\Gamma_{[a_1a_2}{\cal C})^{\alpha(\beta}(\Gamma_{a_3a_4]}{\cal C})^{\gamma\delta)},
\label{fierz3} \\
0 & = & (\Gamma_{c_1c_2}{\cal C})^{\alpha(\beta}(\Gamma^{ba_1a_2a_3c_1c_2}{\cal C})^{\gamma\delta)}
 -(\Gamma^{ba_1a_2a_3c_1c_2}{\cal C})^{\alpha(\beta}(\Gamma_{c_1c_2}{\cal C})^{\gamma\delta)}
\nn & &
 +2(\Gamma_{c}{\cal C})^{\alpha(\beta}(\Gamma^{ba_1a_2a_3c}{\cal C})^{\gamma\delta)}
 -2(\Gamma^{ba_1a_2a_3c}{\cal C})^{\alpha(\beta}(\Gamma_{c}{\cal C})^{\gamma\delta)}
\nn & &
 +8(\Gamma^{a_1a_2a_3}{\cal C})^{\alpha(\beta}(\Gamma^{b}{\cal C})^{\gamma\delta)}
 -24(\Gamma^{b[a_1a_2}{\cal C})^{\alpha(\beta}(\Gamma^{a_3]}{\cal C})^{\gamma\delta)}.
\label{fierz4} 
\eea
The first one in the above is well-known: it ensures M2-brane kappa symmetry.
However, rather than to apply \eqref{ft3}, \eqref{ft4}, and \eqref{ft5} directly,
it is easier to apply the following procedure:
From \eqref{ft3} we obtain
\bea
(\Gamma^b{\cal C})^{\alpha(\beta}(\Gamma_{ab}{\cal C})^{\gamma\delta)}
 & = & -(\Gamma_a\Gamma^b{\cal C})^{\alpha(\beta}(\Gamma_b{\cal C})^{\gamma\delta)}
 +({\cal C})^{\alpha(\beta}(\Gamma_a{\cal C})^{\gamma\delta)},
\label{redft1} \\
(\Gamma^b{\cal C})^{\alpha(\beta}(\Gamma_{a_1\dots a_4b}{\cal C})^{\gamma\delta)}
 & = & 6(\Gamma_{[a_1a_2}{\cal C})^{\alpha(\beta}(\Gamma_{a_3a_4]}{\cal C})^{\gamma\delta)}
 -(\Gamma_{a_1\dots a_4}\Gamma^b{\cal C})^{\alpha(\beta}(\Gamma_b{\cal C})^{\gamma\delta)}
\nn & &
 +4(\Gamma_{[a_1a_2a_3}{\cal C})^{\alpha(\beta}(\Gamma_{a_4]}{\cal C})^{\gamma\delta)}.
\label{redft2}
\eea
Note that in the above relations the number of local Lorentz indices of gamma matrices
with spinor indices symmetrized on the right hand sides are smaller than those on the left hand sides. 
Therefore we can use them for reducing the numbers of local Lorentz indices of
gamma matrices sandwiched by $c$, 
if they are equal to 2 or 5 and some of the indices are contracted.
If the gamma matrices have more than 5 indices, we can reduce the number 
by the following double duality relation without totally antisymmetric tensors:
For $n\le m$, 
\begin{multline}
(\Gamma_{a_1\dots a_n}{}^{c_1\dots c_l})^\alpha{}_\beta
(\Gamma^{b_1\dots b_m}{}_{c_1\dots c_l})^\gamma{}_\delta
\\ 
 = (-1)^{1+n(n-1)/2+m(m-1)/2}
\frac{l!(11-l)!}{(11-n-l)!(11-m-l)!}
\\ \times
 \delta_{a_1}{}^{[b_1}\dots\delta_{a_n}{}^{b_n}
 \delta_{d_1}{}^{b_{n+1}}\dots\delta_{d_{m-n}}{}^{b_m}\delta_{d_{m-n+1}}{}^{e_1}\dots
 \delta_{d_{11-n-l}}{}^{e_{11-m-l}]}
\\ \times
 (\Gamma^{d_1\dots d_{11-n-l}})^\alpha{}_\beta
 (\Gamma_{e_1\dots e_{11-m-l}})^\gamma{}_\delta.
\label{appa:2bldualized}
\end{multline}
By applying these repeatedly for reducing the numbers of indices as much as possible,
we see the cancellation of terms more easily.
In Mathematica calculations, especially for terms in the form of
$\bar{c}\Gamma_1c\cdot\bar{c}\Gamma_2c\cdot\bar{c}\Gamma_3$,
this procedure gives an algorithm much faster than using
\eqref{ft3}, \eqref{ft4}, and \eqref{ft5} directly.

\eqref{appa:2bldualized} also means that there exist
some relations between $\bar{c}\Gamma_{a_1\dots a_n}{}^{c_1\dots c_l}c
\bar{c}\Gamma^{b_1\dots b_m}{}_{c_1\dots c_l}c$ for $n+m+2l=11$.
They are given by
\bea
0 & = & \bar{c}\Gamma_{c_1\dots c_5}c\bar{c}\Gamma^{bc_1\dots c_5}c,
\nn
0 & = & \bar{c}\Gamma_{a_1a_2c_1\dots c_3}c\bar{c}\Gamma^{b_1b_2b_3c_1\dots c_3}c
 +\frac{3}{4}\delta_{[a_1}{}^{[b_1}\bar{c}\Gamma_{a_2]c_1\dots c_4}c\bar{c}\Gamma^{b_1b_2]c_1\dots c_4}c,
\nn
0 & = & \bar{c}\Gamma_{a_1a_2a_3a_4c_1}c\bar{c}\Gamma^{b_1b_2b_3b_4b_5c_1}c
 +5\delta_{[a_1}{}^{[b_1}\bar{c}\Gamma_{a_2a_3a_4]c_1c_2}c\bar{c}\Gamma^{b_2b_3b_4b_5]c_1c_2}c
\nn & &
 -\frac{5}{2}\delta_{[a_1}{}^{[b_1}\delta_{a_2}{}^{b_2}\delta_{a_3}{}^{b_3}
  \bar{c}\Gamma_{a_4]c_1\dots c_4}c\bar{c}\Gamma^{b_4b_5]c_1\dots c_4}c,
\nn
0 & = & \bar{c}\Gamma_{c}c\bar{c}\Gamma^{b_1\dots b_9c}c,
\nn
0 & = & \bar{c}\Gamma_{a_1c}c\bar{c}\Gamma^{b_1\dots b_8c}c
 +2\delta_{a_1}{}^{[b_1}\bar{c}\Gamma_{c_1c_2}c\bar{c}\Gamma^{b_2\dots b_8]c_1c_2}c.
\eea
These also help us to see the cancellation of terms in the form of
$\bar{c}\Gamma_1c\cdot\bar{c}\Gamma_2c$.

\section{Details of $A$ and $B$}
\label{appb}

Let us calculate 
\bea
A^{A_0} & = & (-1)^{C_0}A^{A_0}{}_{B_0C_0D_0}C^{D_0}C^{C_0}C^{B_0},
\\
B^{B_{-1}A_{-1}} & = & (-1)^{C_0}B^{B_{-1}A_{-1}}{}_{B_0C_0D_0}C^{D_0}C^{C_0}C^{B_0},
\eea
by setting the symbols in the definition \eqref{defA} and \eqref{defB} to those given by
\eqref{defTA}, \eqref{defTL}, \eqref{defTD}, \eqref{defTS} and \eqref{defE},
and see if they are indeed in the form of \eqref{reducedA} and \eqref{reducedB}.

First let us calculate $A^{A_0}$.
Most of the calculation is straightforward, except that
we need Fierz identity \eqref{fierz1} for $A_0=(A)$ and \eqref{fierz3} for $A_0=(S)$.
The result is 
\bea
A^{[\mu\nu]} & = & R^{[\mu\nu]}{}_{A_1}F^{A_1} = \p_{\mu}F^{[\nu]}-\p_{\nu}F^{[\mu]},
\\
A^{[ab]} & = & \p{\cal S}_0/\p\psi_\mu^\alpha D^{(\mu\alpha)[ab]}
 =-ieD^{(\mu\alpha)[ab]}({\cal C}^{-1}\Gamma^{\mu\nu\lambda}\wt{D}_\nu\psi_\lambda)_\alpha,
\\
A^{(D)} & = & 0,
\\
A^{(S)} & = & 0,
\eea
where
\beq
F^{[\mu]} = 3c^\nu c^\lambda \p_{[\mu}c_{\nu\lambda]}
 -\frac{i}{4}c^\nu\bar{c}\Gamma^\lambda c A_{\mu\nu\lambda}
 -\frac{i}{4}c^\nu\bar{c}\Gamma_{\mu\nu}c,
\label{defF}
\eeq
\bea
D^{(\mu\alpha)[ab]} & = & -\frac{i}{12}\frac{1}{576}e^{-1}\Big[
 \frac{17}{2}\bar{c}\Gamma^c c\bar{c}\Gamma^{ab}{}_{c\mu}{\cal C}
 -\frac{61}{2}\bar{c}\Gamma_\mu c\bar{c}\Gamma^{ab}{\cal C}
 +\bar{c}\Gamma^{[a} c\bar{c}\Gamma^{b]}{}_\mu {\cal C}
\nn & &
 -31e_{\mu}{}^{[a}\bar{c}\Gamma^{b]}c\bar{c}{\cal C}
 -31e_{\mu}{}^{[a}\bar{c}\Gamma_c c\bar{c}\Gamma^{b]c}{\cal C}
\nn & &
 -\frac{11}{4}\bar{c}\Gamma^{c_1c_2}c\bar{c}\Gamma^{ab}{}_{c_1c_2\mu}{\cal C}
 +\frac{7}{2}\bar{c}\Gamma^c{}_\mu c\bar{c}\Gamma^{ab}{}_c{\cal C}
 +\frac{19}{2}\bar{c}\Gamma^{ab}c\bar{c}\Gamma_\mu {\cal C}
\nn & &
 +5\bar{c}\Gamma^{[a}{}_c c\bar{c}\Gamma^{b]c}{}_\mu {\cal C}
 -97\bar{c}\Gamma^{[a}{}_\mu c\bar{c}\Gamma^{b]}{\cal C}
\nn & & 
 -\frac{19}{2}e_{\mu}{}^{[a}\bar{c}\Gamma_{c_1c_2}c\bar{c}\Gamma^{b]c_1c_2}{\cal C}
 +17e_{\mu}{}^{[a}\bar{c}\Gamma^{b]c}c\bar{c}\Gamma_c{\cal C}
\nn & &
 +\frac{1}{24}\bar{c}\Gamma^{[a}{}_{c_1\dots c_4}c\bar{c}\Gamma^{b]c_1\dots c_4}{}_\mu {\cal C}
 -\frac{5}{6}\bar{c}\Gamma^{[a}{}_{c_1c_2c_3\mu}c\bar{c}\Gamma^{b]c_1c_2c_3}{\cal C}
\nn & &
 +\frac{17}{240}\bar{c}\Gamma^{c_1\dots c_5}c\bar{c}\Gamma^{ab}{}_{c_1\dots c_5\mu}{\cal C}
 -\frac{13}{48}\bar{c}\Gamma^{c_1\dots c_4}{}_\mu c\bar{c}\Gamma^{ab}{}_{c_1\dots c_4}{\cal C}
\nn & &
 -\frac{11}{4}\bar{c}\Gamma^{abc_1c_2}{}_\mu c\bar{c}\Gamma_{c_1c_2}{\cal C}
 +\frac{7}{12}\bar{c}\Gamma^{abc_1c_2c_3}c\bar{c}\Gamma_{c_1c_2c_3\mu}{\cal C}
\nn & & 
 -\frac{7}{120}e_{\mu}{}^{[a}\bar{c}\Gamma^{c_1\dots c_5}c\bar{c}\Gamma^{b]}{}_{c_1\dots c_5}{\cal C}
 -\frac{7}{24}e_{\mu}{}^{[a}\bar{c}\Gamma^{b]c_1\dots c_4}c\bar{c}\Gamma_{c_1\dots c_4}{\cal C}
\Big]^\alpha.
\label{defD}
\eea
The ambiguity in the definition of $F^{A_1}$ is fixed so that
$F^{A_1}$ contains $c_{\mu\nu}$ only in the form of its field strength, and
the ambiguity in the definition of $D^{(\mu\alpha)[ab]}$ is fixed similarly to $E^{B_{-1}A_{-1}}{}_{B_0C_0}$.
The expression \eqref{defD} is already symmetrized under interchange of three $c^\epsilon$s.
i.e. $D^{(\mu\alpha)[ab]}{}_{(\beta)(\gamma)(\delta)}$ is given just by removing $c^\epsilon$
in \eqref{defD}, and putting indices $\beta, \gamma$ and $\delta$: 
\bea
D^{(\mu\alpha)[ab]}{}_{(\beta)(\gamma)(\delta)} & = & -\frac{i}{12}\frac{1}{576}e^{-1}\Big[
 \frac{17}{2}({\cal C}^{-1}\Gamma^c)_{\beta\gamma}({\cal C}^{-1}\Gamma^{ab}{}_{c\mu}{\cal C})_{\delta}{}^\alpha
\nn & & 
 -\frac{61}{2}({\cal C}^{-1}\Gamma_\mu)_{\beta\gamma}({\cal C}^{-1}\Gamma^{ab}{\cal C})_{\delta}{}^\alpha
 +\dots \Big].
\label{redD}
\eea
This can be seen from the fact that \eqref{defD} is invariant if we apply \eqref{ft3}.
However the following reduced expression, which is given by applying \eqref{redft1}, \eqref{redft2}, and
\eqref{appa:2bldualized} to \eqref{defD}, is simpler:
\bea
D^{(\mu\alpha)[ab]} & = & -\frac{i}{288}e^{-1}\Big[
 -6e_{\mu}{}^{[a}\bar{c}\Gamma_c c\bar{c}\Gamma^{b]c}{\cal C}
 -3\bar{c}\Gamma_\mu c\bar{c}\Gamma^{ab}{\cal C}
\nn & &
 +2\bar{c}\Gamma^{ab} c\bar{c}\Gamma_\mu {\cal C}
 +12\bar{c}\Gamma_\mu{}^{[a}c\bar{c}\Gamma^{b]}{\cal C}
\Big]^\alpha.
\label{defD2}
\eea
If we read off $D^{(\mu\alpha)[ab]}$ from this expression we need explicit symmetrization of indices:
\bea
D^{(\mu\alpha)[ab]}{}_{(\beta)(\gamma)(\delta)} & = & -\frac{i}{288}e^{-1}\Big[
 -6e_{\mu}{}^{[a}({\cal C}^{-1}\Gamma_c)_{(\beta\gamma}({\cal C}^{-1}\Gamma^{b]c}{\cal C})_{\delta)}{}^\alpha
\nn & &
 -3({\cal C}^{-1}\Gamma_\mu)_{(\beta\gamma}({\cal C}^{-1}\Gamma^{ab}{\cal C})_{\delta)}{}^\alpha+\dots\Big].
\eea
Thus we see that $D^{(\mu\alpha)[ab]}$ is the only nonzero component of $D^{B_{-1}A_{-1}}$.

It is an important check to confirm that we can obtain the same 
$D^{B_{-1}A_{-1}}$ from $B^{B_{-1}A_{-1}}$. 
From \eqref{defB} and \eqref{Econd},
we see that $B^{B_{-1}A_{-1}}$ does not vanish only if either $B_{-1}$ or $A_{-1}$ is $(\psi)$,
and
\bea
B^{(\nu a)(\mu\alpha)} & = & -\frac{i}{4}(\bar{c}\Gamma^a)_\beta E^{(\nu\beta)(\mu\alpha)},
\\
B^{(\rho\alpha)[\mu\nu\lambda]}
 & = & -\frac{3}{4}i(\bar{c}\Gamma_{[\mu\nu})_\beta E^{(\rho\alpha)(\lambda]\beta)}.
\label{BpsiA}
\eea
Using \eqref{defE}, both of the above are expressed
by sums of terms in the form of $\bar{c}\Gamma_1c(\bar{c}\Gamma_2{\cal C})^\alpha$.
By performing Fierz transformation \eqref{ft3} (or the faster procedure) to them, we see that
$B^{(\rho\alpha)[\mu\nu\lambda]}$ vanishes, and $B^{(\nu a)(\mu\alpha)}$
indeed gives the same $D^{(\mu\alpha)[ab]}$ as \eqref{defD}:
\bea
B^{(\nu a)(\mu\alpha)} & = & e_{\nu b}D^{(\mu\alpha)[ab]},
\\
B^{(\rho\alpha)[\mu\nu\lambda]} & = & 0.
\eea
Then the final task is to calculate $B^{(\nu\beta)(\mu\alpha)}$.
Because we need similar calculations in the following appendices, 
we explain the detail of the calculation in this case.
From \eqref{defB} and \eqref{Econd},
\begin{multline}
B^{(\nu\beta)(\mu\alpha)} = 
 \p E^{(\nu\beta)(\mu\alpha)}/\p e_\lambda{}^a R^{(\lambda a)}{}_{A_0}C^{A_0}
 -E^{(\nu\beta)(\mu\alpha)}{}_{(\gamma)(\delta)}T^{(\delta)}c^\gamma
\\
 +\p(R^{(\nu\beta)}{}_{A_0}C^{A_0})/\p\psi_\lambda^\delta E^{(\lambda\delta)(\mu\alpha)}
 +\p(R^{(\mu\alpha)}{}_{A_0}C^{A_0})/\p\psi_\lambda^\delta E^{(\lambda\delta)(\nu\beta)}.
\label{Bpsipsi}
\end{multline}
The first term in \eqref{Bpsipsi} can be calculated by making the symmetry transformation for $e_\mu{}^a$
in $E^{(\nu\beta)(\mu\alpha)}$ with transformation parameters replaced by $C^{A_0}$,
which is denoted by $\wt{\delta}$.
This replacement is done after reordering the parameters to the rightmost position:
\bea
\wt{\delta}e_\mu{}^a & = &
 (\wt{\delta}^S+\wt{\delta}^D+\wt{\delta}^L+\wt{\delta}^A)e_\mu{}^a
\nn & = & -\frac{i}{4}\bar{c}\Gamma^a\psi_\mu
-c^\nu\p_\nu e_\mu{}^a-\p_\mu c^\nu e_\nu{}^a
+c^a{}_be_\mu{}^b,
\eea
where the first term obtains an additional sign factor due to the reordering.
$\wt{\delta}\psi_\mu^\alpha$ is defined similarly.

The second term in \eqref{Bpsipsi} can be calculated just by replacing one of $c^\gamma$s in
$E^{(\nu\beta)(\mu\alpha)}/2$ by $T^{(\gamma)}$:
\begin{multline}
-E^{(\nu\beta)(\mu\alpha)}{}_{(\gamma)(\delta)}T^{(\delta)}c^\gamma
=\frac{1}{2}\p E^{(\nu\beta)(\mu\alpha)}/\p c^\gamma T^{(\gamma)}
\\
=\p E^{(\nu\beta)(\mu\alpha)}/\p c^\gamma (\frac{1}{4}c_{ab}\Gamma^{ab}c)^\gamma
 -c^\lambda\p E^{(\nu\beta)(\mu\alpha)}/\p c^\gamma \p_\lambda c^\gamma
 -\frac{i}{8}\bar{c}\Gamma^\lambda c \p E^{(\nu\beta)(\mu\alpha)}/\p c^\gamma\psi_\lambda^\gamma.
\end{multline}
The third and fourth terms in \eqref{Bpsipsi} are rewritten as
\bea
\p(\wt{\delta}\psi_\nu^\beta)/\p\psi_\lambda^\delta E^{(\lambda\delta)(\mu\alpha)}
 +((\mu\alpha)\leftrightarrow (\nu\beta)).
\eea
Let us calculate this for each of $\wt{\delta}^S, \wt{\delta}^D$, and $\wt{\delta}^L$ in $\wt{\delta}$:
\begin{multline}
\p(\wt{\delta}^L\psi_\nu^\beta)/\p\psi_\lambda^\delta E^{(\lambda\delta)(\mu\alpha)}
 +((\mu\alpha)\leftrightarrow (\nu\beta))
\\
=-\frac{1}{4}c_{ab}(\Gamma^{ab})^\beta{}_\gamma E^{(\nu\gamma)(\mu\alpha)}
 -\frac{1}{4}c_{ab}(\Gamma^{ab})^\alpha{}_\gamma E^{(\mu\gamma)(\nu\beta)},
\end{multline}
\begin{multline}
\p(\wt{\delta}^S\psi_\nu^\beta)/\p\psi_\lambda^\delta E^{(\lambda\delta)(\mu\alpha)}
 +((\mu\alpha)\leftrightarrow (\nu\beta))
\\
=\frac{1}{4}(\Gamma^{bc}c)^\beta
 \p\hat{\omega}_{\nu ab}/\p\psi_\lambda^\gamma E^{(\lambda\gamma)(\mu\alpha)}
+\frac{1}{288}[(\Gamma^{\mu_1\dots\mu_4}{}_\nu
 -8\delta_\nu{}^{\mu_1}\Gamma^{\mu_2\mu_3\mu_4})c]^\beta
 \p\hat{F}_{\mu_1\dots \mu_4}/\p\psi_\lambda^\gamma E^{(\lambda\gamma)(\mu\alpha)}
\\
+((\mu\alpha)\leftrightarrow (\nu\beta)),
\end{multline}
and we have to be careful with $\wt{\delta}^D$ because it contains derivative operators.
Indices $(\mu\alpha)$ and $(\nu\beta)$ must be supplemented with spacetime positions:
$(\mu\alpha)\rightarrow(\mu\alpha x)$ and $(\nu\beta)\rightarrow(\nu\beta y)$.
Then,
\begin{multline}
\p(\wt{\delta}^D\psi_\nu^\beta)/\p\psi_\lambda^\delta E^{(\lambda\delta)(\mu\alpha)}
 +((\mu\alpha)\leftrightarrow (\nu\beta))
\\
=\p_\nu c^\lambda(y) E^{(\lambda\beta)(\mu\alpha)}(y)\delta(y-x)
 +\int dz c^\lambda(y)\p_\lambda^y\delta(y-z) E^{(\nu\beta)(\mu\alpha)}(z)\delta(z-x)
\\
 +((\mu\alpha x)\leftrightarrow (\nu\beta y))
\\
=(\p_\nu c^\lambda(y) E^{(\lambda\beta)(\mu\alpha)}(y)
 +\p_\mu c^\lambda(y) E^{(\nu\beta)(\lambda\alpha)}(y))\delta(y-x)
\\
+(c^\lambda(y)E^{(\nu\beta)(\mu\alpha)}(x)
 -c^\lambda(x)E^{(\nu\beta)(\mu\alpha)}(y))\p_\lambda^y\delta(y-x)
\\
=(\p_\nu c^\lambda(y) E^{(\lambda\beta)(\mu\alpha)}(y)
 +\p_\mu c^\lambda(y) E^{(\nu\beta)(\lambda\alpha)}(y)
\\
 -\p_\lambda c^\lambda(y) E^{(\nu\beta)(\mu\alpha)}(y)
 +c^\lambda(y)\p_\lambda E^{(\nu\beta)(\mu\alpha)}(y)
)\delta(y-x).
\label{Bpsipsi2}
\end{multline}
Noting that $E^{(\nu\beta)(\mu\alpha)}$ is in the following form,
\beq
E^{(\nu\beta)(\mu\alpha)}=e^{-1}e_\mu{}^a e_\nu{}^b\cdot\text{($e_\lambda{}^c$-independent part)},
\eeq
\eqref{Bpsipsi2} is equal to minus the diffeomorphism transformation of $E^{(\nu\beta)(\mu\alpha)}$
with the parameter replaced by $c^\lambda$.
and it is equal to 
\beq
-\wt{\delta}^DE^{(\nu\beta)(\mu\alpha)}
+c^\lambda\p E^{(\nu\beta)(\mu\alpha)}/\p c^\gamma \p_\lambda c^\gamma.
\eeq
(Note that $\wt{\delta}^D$ does not act on $c^\gamma$.)
Hence terms proportional to $c^\lambda$ in \eqref{Bpsipsi} cancel.

Next let us collect terms proportional to $c_{ab}$ in \eqref{Bpsipsi}:
\bea
\wt{\delta}^LE^{(\nu\beta)(\mu\alpha)}
+\p E^{(\nu\beta)(\mu\alpha)}/\p c^\gamma (\frac{1}{4}c_{ab}\Gamma^{ab}c)^\gamma
\nn
-\frac{1}{4}c_{ab}(\Gamma^{ab})^\beta{}_\gamma E^{(\nu\gamma)(\mu\alpha)}
-\frac{1}{4}c_{ab}(\Gamma^{ab})^\alpha{}_\gamma E^{(\nu\beta)(\mu\gamma)}.
\eea
We see that the above vanish again, because the first and second term give local Lorentz
transformation of $E^{(\nu\beta)(\mu\alpha)}$ with the parameter replaced by $c_{ab}$,
which cancels the third and fourth term.

Then the remaining terms in \eqref{Bpsipsi} are given by
\bea
B^{(\nu\beta)(\mu\alpha)} =
\wt{\delta}^S E^{(\nu\beta)(\mu\alpha)}
-\frac{i}{8}\bar{c}\Gamma^\lambda c \p E^{(\nu\beta)(\mu\alpha)}/\p c^\gamma\psi_\lambda^\gamma
+\frac{1}{4}(\Gamma^{bc}c)^\beta
 \p\hat{\omega}_{\nu ab}/\p\psi_\lambda^\gamma E^{(\lambda\gamma)(\mu\alpha)}
\nn
+\frac{1}{288}[(\Gamma^{\mu_1\dots\mu_4}{}_\nu
 -8\delta_\nu{}^{\mu_1}\Gamma^{\mu_2\mu_3\mu_4})c]^\beta
 \p\hat{F}_{\mu_1\dots \mu_4}/\p\psi_\lambda^\gamma E^{(\lambda\gamma)(\mu\alpha)}.
\label{Bfinal1}
\eea
This does not contain derivative operators, and therefore 
this cannot have terms proportional to the equation of motion. Hence
\beq
M^{C_{-1}B_{-1}A_{-1}}{}_{B_0C_0D_0}=0.
\eeq
Then from \eqref{reducedB}, $B^{(\nu\beta)(\mu\alpha)}$ must be in the following form:
\bea
B^{(\nu\beta)(\mu\alpha)} & = &
 R^{(\nu\beta)}{}_{[ab]}D^{(\mu\alpha)[ab]}
 +R^{(\mu\alpha)}{}_{[ab]}D^{(\nu\beta)[ab]}
\nn & = &
 \frac{1}{4}(\Gamma_{ab}\psi_\nu)^\beta D^{(\mu\alpha)[ab]}
 +\frac{1}{4}(\Gamma_{ab}\psi_\mu)^\alpha D^{(\nu\beta)[ab]}.
\label{Bfinal2}
\eea
To confirm that \eqref{Bfinal1} is indeed equal to \eqref{Bfinal2},
we need Fierz transformation: both expressions consist of 
terms containing three $c^\gamma$s and one $\psi_\lambda$.
They can be rearranged to the form 
$\bar{c}\Gamma_1c\cdot\bar{c}\Gamma_2\psi_\lambda\cdot
 (\Gamma_{a_1\dots a_n}{\cal C})^{\alpha\beta}~(n=0,1,\dots,5)$.
The coefficients of $(\Gamma_{a_1\dots a_n}{\cal C})^{\alpha\beta}$ can be
obtained by multiplying $({\cal C}^{-1}\Gamma^{a_1\dots a_n})_{\beta\alpha}$
to \eqref{Bfinal1} or \eqref{Bfinal2}.
Applying \eqref{ft3} (or the faster procedure) to those coefficients we see that the difference between
\eqref{Bfinal1} and \eqref{Bfinal2} vanishes.

\section{Details of $Z$ and $W$}
\label{appc}

Let us calculate 
\bea
Z^{A_1} & = &
 (-1)^{B_0+D_0}Z^{A_1}{}_{B_0C_0D_0E_0}C^{E_0}C^{D_0}C^{C_0}C^{B_0},
\\
W^{B_{-1}A_0} & = & 
 (-1)^{B_0+D_0}W^{B_{-1}A_0}{}_{B_0C_0D_0E_0}
 C^{E_0}C^{D_0}C^{C_0}C^{B_0},
\eea
by setting the symbols in the definition \eqref{defZ} and \eqref{defW} to those given by
\eqref{defTA}-\eqref{defTS}, \eqref{defE}, and \eqref{defD} (or \eqref{redD}).

Calculation of $Z^{A_1}$ is straightforward, except that we need Fierz identity 
\eqref{fierz1} to cancel terms proportional to $\psi_\mu^\alpha$. The result is
\beq
Z^{A_1} = R^{A_1}{}_{A_2}Y^{A_2}=\p_\mu Y,
\eeq
\beq
Y=-c^\mu c^\nu c^\lambda \p_{[\mu}c_{\nu\lambda]}
 +\frac{i}{8}c^\mu c^\nu\bar{c}\Gamma^\lambda cA_{\mu\nu\lambda}
 +\frac{i}{8}c^\mu c^\nu\bar{c}\Gamma_{\mu\nu} c.
\label{defY}
\eeq
From \eqref{Econd} and \eqref{Dcond} we can easily see that
some components of $W^{B_{-1}A_0}$ vanish. Especially
for $A_0=(A),(D)$, and $(S)$, $B_{-1}$ must be $(\psi)$ to give nonzero contribution.
Then because $T^{(A)}{}_{B_0C_0}$ and $T^{(D)}{}_{B_0C_0}$ vanish if
either $B_0$ or $C_0$ is $(L)$, $W^{B_{-1}A_0}$ vanishes for $A_0=(A)$ or $(D)$.

The remaining nontrivial components are $W^{(\mu\alpha)[ab]}$ and
\bea
W^{[\mu\nu\lambda][ab]} & = &
 -\frac{3}{4}i(\bar{c}\Gamma_{[\mu\nu})_\alpha D^{(\lambda]\alpha)[ab]}
\nn & = & -\frac{3}{16}i(\bar{c}\Gamma_{[\mu\nu})_\alpha 
 (\bar{c}\Gamma^{[a})_\beta e^{b]\rho}E^{(\lambda]\alpha)(\rho\beta)},
\\
W^{(\mu c)[ab]} & = &
 \frac{i}{4}(\bar{c}\Gamma^c)_\alpha D^{(\mu\alpha)[ab]},
\\
W^{(\mu\alpha)(\beta)} & = &
 \frac{1}{4}(\Gamma^{ab}c)^\beta D^{(\mu\alpha)[ab]}
 +\frac{i}{8}(\bar{c}\Gamma^\nu c)E^{(\nu\beta)(\mu\alpha)}.
\eea
$W^{[\mu\nu\lambda][ab]}$ is proportional to \eqref{BpsiA}, and therefore
vanishes. We can see that $W^{(\mu\alpha)(\beta)}$ also vanishes:
By using \eqref{defE} and \eqref{defD} or \eqref{redD}, and rearranging the resulting terms,
$W^{(\mu\alpha)(\beta)}$ is expressed by a sum of terms in the form of 
$\bar{c}\Gamma_1c\cdot\bar{c}\Gamma_2c\cdot
(\Gamma_{a_1\dots a_n}{\cal C})^{\alpha\beta}~(n=0,1,\dots,5)$.
The coefficients of $(\Gamma_{a_1\dots a_n}{\cal C})^{\alpha\beta}$ can be obtained by
multiplying $({\cal C}^{-1}\Gamma^{a_1\dots a_n})_{\alpha\beta}$ to 
$W^{(\mu\alpha)(\beta)}$. Performing Fierz transformation \eqref{ft4} (or the faster procedure)
to the coefficients we see them vanish.

It is not difficult to see that $W^{(\mu c)[ab]}$ is in the following form: 
\beq
W^{(\mu c)[ab]} = R^{(\mu c)}{}_{[de]}V^{[de][ab]}
 = e_\mu{}^dV^{[cd][ab]},
\eeq
where
\bea
V^{[cd][ab]} & = & -\frac{1}{48}\frac{1}{576}e^{-1}\Big[
 -31\bar{c}\Gamma^{a_1}c\bar{c}\Gamma_{a_1}c\delta_{[c}{}^a\delta_{d]}{}^b
\nn & & 
 +\frac{19}{2}\bar{c}\Gamma^{a_1a_2}c\bar{c}\Gamma_{a_1a_2}c\delta_{[c}{}^a\delta_{d]}{}^b
 -\frac{7}{120}\bar{c}\Gamma^{a_1\dots a_5}c\bar{c}\Gamma_{a_1\dots a_5}c\delta_{[c}{}^a\delta_{d]}{}^b
\nn & & 
 -6\bar{c}\Gamma_{cd}c\bar{c}\Gamma^{ab}c
 +92\bar{c}\Gamma_{[c}{}^ac\bar{c}\Gamma_{d]}{}^bc
 +\frac{2}{3}\bar{c}\Gamma_{[c}{}^{a_1a_2a_3a}c\bar{c}\Gamma_{d]a_1a_2a_3}{}^bc
\nn & & 
 +124\bar{c}\Gamma^{[a}c\bar{c}\Gamma_{[c}c\delta_{d]}{}^{b]}
 -14\bar{c}\Gamma^{a_1}c\bar{c}\Gamma_{a_1cd}{}^{ab}c
\nn & & 
 -4\bar{c}\Gamma^{a_1[a}c\bar{c}\Gamma_{a_1[c}c\delta_{d]}{}^{b]}
 +7\bar{c}\Gamma^{a_1a_2}c\bar{c}\Gamma_{a_1a_2cd}{}^{ab}c
\nn & & 
 -\frac{5}{3}\bar{c}\Gamma_{a_1a_2a_3cd}c\bar{c}\Gamma^{a_1a_2a_3ab}c
 +\frac{7}{6}\bar{c}\Gamma^{a_1\dots a_4[a}c\bar{c}\Gamma_{a_1\dots a_4[c}c\delta_{d]}{}^{b]}
\Big].
\label{defV}
\eea
Note that $V^{[cd][ab]}=V^{[ab][cd]}$. This expression is already symmetrized 
under interchange of four $c^\alpha$s. It can be shown by seeing that \eqref{defV} is invariant
if we apply \eqref{ft4}. However the following reduced form
given by applying \eqref{redft1}, \eqref{redft2}, and
\eqref{appa:2bldualized} to \eqref{defV} is simpler:
\bea
V^{[cd][ab]} & = & \frac{1}{576}e^{-1}\Big[
 3\bar{c}\Gamma^{a_1}c\bar{c}\Gamma_{a_1}c\delta_{[c}{}^a\delta_{d]}{}^b
 -6\bar{c}\Gamma^{[a}c\bar{c}\Gamma_{[c}c\delta_{d]}{}^{b]}
\nn & & 
 +\bar{c}\Gamma^{ab}c\bar{c}\Gamma_{cd}c
 -6\bar{c}\Gamma_{[c}{}^ac\bar{c}\Gamma_{d]}{}^bc
\Big].
\label{redV}
\eea
Then we infer that $W^{(\mu\alpha)[ab]}$ is given by
\beq
W^{(\mu\alpha)[ab]} = R^{(\mu\alpha)}{}_{[cd]}V^{[cd][ab]}
 =\frac{1}{4}V^{[ab][cd]}(\Gamma^{cd}\psi_\mu)^\alpha.
\label{reducedW}
\eeq
Indeed this is correct. From \eqref{defW},
\bea
W^{(\mu\alpha)[ab]} & = & 
-\wt{\delta}D^{(\mu\alpha)[ab]}
-\frac{1}{2}\p D^{(\mu\alpha)[ab]}/\p c^\gamma T^{(\gamma)}
\nn & &
-\frac{1}{2}\p T^{[ab]}/\p c^{[cd]} D^{(\mu\alpha)[cd]}
-\p\wt{\delta}\psi_\mu^\alpha/\p\psi_\nu^\beta D^{(\nu\beta)[ab]}
\nn & &
+\frac{1}{2}\p T^{[ab]}/\p\psi_\nu^\beta E^{(\mu\alpha)(\nu\beta)},
\eea
where we made a manipulation similar to $B^{(\nu\beta)(\mu\alpha)}$
in Appendix \ref{appb}. Terms containing $c^{[cd]}$ and $c^\lambda$
cancel again by an argument similar in Appendix \ref{appb}.
Therefore $W^{(\mu\alpha)[ab]}$ contains terms with 4 $c^\gamma$s and one $\psi_\lambda$:
\bea
W^{(\mu\alpha)[ab]} & = & -\wt{\delta}^SD^{(\mu\alpha)[ab]}
+\frac{i}{8}\bar{c}\Gamma^\nu c\p D^{(\mu\alpha)[ab]}/\p c^\gamma \psi_\nu^\gamma
\nn & &
-\frac{1}{4}(\Gamma^{cd}c)^\alpha
 \p\hat{\omega}_{\mu cd}/\p\psi_\nu^\beta D^{(\mu\beta)[ab]}
\nn & &
-\frac{1}{288}[(\Gamma^{\mu_1\dots\mu_4}{}_\mu
 -8\delta_\mu{}^{\mu_1}\Gamma^{\mu_2\mu_3\mu_4})c]^\alpha
 \p\hat{F}_{\mu_1\dots\mu_4}/\p\psi_\nu^\beta D^{(\nu\beta)[ab]}
\nn & & 
+\frac{i}{8}\bar{c}\Gamma^\lambda c
 \p\hat{\omega}_\lambda{}^{ab}/\p\psi_\nu^\beta E^{(\mu\alpha)(\nu\beta)}
\nn & &
+\frac{i}{1152}\bar{c}(\Gamma^{ab\mu_1\dots\mu_4}
 +24e^{a\mu_1}e^{b\mu_2}\Gamma^{\mu_3\mu_4})c
 \p\hat{F}_{\mu_1\dots\mu_4}/\p\psi_\nu^\beta E^{(\mu\alpha)(\nu\beta)}.
\eea
Rearranging terms in the above into the form of 
$\bar{c}\Gamma_1c\cdot\bar{c}\Gamma_2c\cdot \Gamma_{a_1\dots a_n}\psi_\lambda$,
and applying \eqref{ft4} (or the faster procedure) to the coefficients of 
$\Gamma_{a_1\dots a_n}\psi_\lambda$,
we see that \eqref{reducedW} is correct.

In summary, $W^{B_{-1}A_0}$ is in the following form:
\beq
W^{B_{-1}A_0}=R^{B_{-1}}{}_{B_0}V^{B_0A_0},
\eeq
and the only nonzero component of $V^{B_0A_0}$ is $V^{(L)(L)}$
given by \eqref{defV} or \eqref{redV}.

\section{Details of $X$ and $U$}
\label{appd}

Let us calculate
\bea
X^{A_2} & := & (-1)^{C_0+E_0}X^{A_2}{}_{B_0C_0D_0E_0F_0}
 C^{F_0}C^{E_0}C^{D_0}C^{C_0}C^{B_0},
\\
U^{B_0A_0} & := & (-1)^{D_0+F_0}U^{B_0A_0}{}_{C_0D_0E_0F_0G_0}
 C^{G_0}C^{F_0}C^{E_0}C^{D_0}C^{C_0},
\eea
by setting the symbols in the definition \eqref{defX} and \eqref{defU} to those given by
\eqref{defTA}-\eqref{defTS}, \eqref{defE}, \eqref{defD} (or \eqref{redD}), 
\eqref{defY} and \eqref{defV} (or \eqref{redV}).

It is not difficult to see that by straightforward calculation with \eqref{fierz1},
\beq
X^{A_2}=0.
\eeq

From \eqref{Dcond} and \eqref{Vcond}, we see that some components of $U^{B_0A_0}$ vanish,
and nontrivial components are given for $(A_0,B_0)=(L,A), (L,D), (L,S)$ and $(L,L)$.
Since $T^{[\mu\nu]}$ and $T^{(\mu)}$ do not depend on $\psi_\mu^\alpha$ and $c_{[ab]}$, 
$U^{(L)(A)}$ and $U^{(L)(D)}$ vanish.
For $(A_0,B_0)=(L,S)$,
\beq
U^{[ab](\alpha)}=\frac{1}{4}V^{[ab][cd]}(\Gamma^{cd}c)^\alpha
 -\frac{i}{8}\bar{c}\Gamma^\mu cD^{(\mu\alpha)[ab]}.
\eeq
By using \eqref{defD} and \eqref{defV} (or, \eqref{redD} and \eqref{redV}), we see that
$U^{[ab](\alpha)}$ consists of terms with 5 $c^\gamma$s.
They can be rearranged to terms in the form of 
$(\bar{c}\Gamma_1)^\alpha\cdot\bar{c}\Gamma_2c\cdot\bar{c}\Gamma_3c$.
Applying \eqref{ft5} (or the faster procedure) to those terms we see that $U^{[ab](\alpha)}$ vanishes.

For $(A_0,B_0)=(L,L)$,
\bea
U^{[ab][cd]} & = & \wt{\delta}V^{[cd][ab]}
 +\frac{1}{2}\p V^{[cd][ab]}/\p c^\alpha T^{(\alpha)}
\nn & &
 +\p T^{[cd]}/\p c^{ef} V^{[ef][ab]}
 +\p T^{[ab]}/\p c^{ef} V^{[ef][cd]}
\nn & &
 +\p T^{[cd]}/\p\psi_\mu^\alpha D^{(\mu\alpha)[ab]}
 +\p T^{[ab]}/\p\psi_\mu^\alpha D^{(\mu\alpha)[cd]},
\eea
where we made a manipulation similar to $B^{(\nu\beta)(\mu\alpha)}$
in Appendix \ref{appb}. Terms containing $c^{[ef]}$ and $c^\mu$
cancel again by an argument similar in Appendix \ref{appb}.
Then,
\bea
U^{[ab][cd]} & = & \wt{\delta}^SV^{[cd][ab]}
 -\frac{i}{8}\bar{c}\Gamma^\mu c\p V^{[cd][ab]}/\p c^\alpha\psi_\mu^\alpha
\nn & &
 +\frac{i}{8}\bar{c}\Gamma^\nu c\p\hat{\omega}_\nu{}^{cd}/\p\psi_\mu^\alpha D^{(\mu\alpha)[ab]}
 +\frac{i}{8}\bar{c}\Gamma^\nu c\p\hat{\omega}_\nu{}^{ab}/\p\psi_\mu^\alpha D^{(\mu\alpha)[cd]}
\nn & &
 +\frac{i}{1152}\bar{c}(\Gamma^{cd\mu_1\dots\mu_4}+24e^{c\mu_1}e^{d\mu_2\Gamma^{\mu_3\mu_4}})c
  \p\hat{F}_{\mu_1\dots\mu_4}/\p\psi_\mu^\alpha D^{(\mu\alpha)[ab]}
\nn & &
 +\frac{i}{1152}\bar{c}(\Gamma^{ab\mu_1\dots\mu_4}+24e^{a\mu_1}e^{b\mu_2\Gamma^{\mu_3\mu_4}})c
  \p\hat{F}_{\mu_1\dots\mu_4}/\p\psi_\mu^\alpha D^{(\mu\alpha)[cd]}.
\eea
We see that $U^{[ab][cd]}$ consists of 
terms in the form of $\bar{c}\Gamma_1 c\cdot\bar{c}\Gamma_2 c\cdot\bar{c}\Gamma_3\psi_\mu$.
By applying \eqref{ft5} (or the faster procedure), we see them cancel.
In summary all the components of $U^{B_0A_0}$ vanish.

\newcommand{\J}[4]{{\sl #1} {\bf #2} (#3) #4}
\newcommand{\andJ}[3]{{\bf #1} (#2) #3}
\newcommand{\AP}{Ann.\ Phys.\ (N.Y.)}
\newcommand{\MPL}{Mod.\ Phys.\ Lett.}
\newcommand{\NP}{Nucl.\ Phys.}
\newcommand{\PL}{Phys.\ Lett.}
\newcommand{\PR}{Phys.\ Rev.}
\newcommand{\PRL}{Phys.\ Rev.\ Lett.}
\newcommand{\PTP}{Prog.\ Theor.\ Phys.}
\newcommand{\hepth}[1]{{\tt hep-th/#1}}

\end{document}